\documentclass[11pt,a4paper]{article}
\usepackage{jheppub}
\usepackage{braket}
\usepackage{tikz}
\usepackage{mathrsfs}  
\usepackage{extarrows}

\begin{document}
\title{\huge Soft theorems in maximally supersymmetric theories}
\author{Zheng-Wen Liu}

\affiliation[]{Department of Physics, \\
Renmin University of China, Beijing 100872, P. R. China}

\emailAdd{zhengwen@ruc.edu.cn}

\abstract{
In this paper we study the supersymmetric generalization of the new soft theorem which was proposed by Cachazo and Strominger recently.
At tree level, we prove the validity of the super soft theorems in both ${\cal N}=4$ super-Yang-Mills theory and ${\cal N}=8$ supergravity using super-BCFW recursion relations.
We verify these theorems exactly by showing some examples.
}
\arxivnumber{1410.1616}
\maketitle

\section{Introduction}
Over the past few decades, there was great progress on the understanding of physical and mathematical structures of the perturbative scattering amplitudes in gauge and gravity theories.
Among these remarkable developments, a recent advance is the soft behavior of scattering amplitudes when one or more external legs tend to zero in both gravity and Yang-Mills theory \cite{Casali-1404.5551,Cachazo-Yuan-1405.3413,
He-Huang-Wen-1405.1410,
Bianchi-He-Huang-Wen-1406.5155,
Bern-Davies-Nohle-1405.1015,
Bern-Davies-Vecchia-Nohle-1406.6987,
Broedel-Leeuw-Plefka-Rosso-1406.6574,
White-1406.7184,Schwab-Volovich-1404.7749,
Afkhami-Jeddi-1405.3533,
Zlotnikov-1407.5936,Kalousios-Rojas-1407.5982,
Schwab-1406.4172,Geyer-Lipstein-Mason-1406.1462,
Larkoski-1405.2346,
Lysov-Pasterski-Strominger-1407.3814,
White-1103.2981,
Adamo-Casali-Skinner-1405.5122,
Strominger-1312.2229,
He-Lysov-Mitra-Strominger-1401.7026,
Kapec-Lysov-Pasterski-Strominger-1406.3312,
He-Mitra-Porfyriadis-Strominger-1407.3789,
Naculich-1407.7836,
Du-Feng-Fu-Wang-1408.4179,
Laenen-Stavenga-White-0811.2067,Laenen-Magnea-Gerben Stavenga-White-1010.1860,Oxburgh-White-1210.1110}.
The study on soft behavior of the amplitudes (in gravity) goes back to Steven Weinberg who proposed the universal leading soft-graviton behavior using the Feynman diagrams technique more than five decades ago \cite{Weinberg1965}.
The leading soft-graviton behavior of the amplitudes, also called ``Weinberg's theorem'', is uncorrected to all loop orders.
At sub-leading order, the soft photon behavior and the soft graviton behavior were also studied by using Feynman diagrams \cite{Low-1954,Low-1958,Gross-Jackiw-1958}.

Recently, sub-leading and sub-sub-leading soft-graviton divergences of amplitudes were proposed \cite{Cachazo-Strominger-1404.4091} beyond Weinberg's theorem.
In \cite{Cachazo-Strominger-1404.4091}, Cachazo and Strominger presented a proof for tree-level amplitudes of gravitons using BCFW recursion relations \cite{Nima-Kaplan-0801.2385,Britto-Cachazo-Feng-0412308,Britto-Cachazo-Feng-Witten-0501052,Cachazo-Svrcek-0502160,Benincasa-Boucher-Veronneau-Cachazo-0702032,Bedford-Brandhuber-Spence-Travaglini-0502146} in the spinor-helicity formulism \cite{Xu-Zhang-Chang-NPB-1987,Dixon-Review-96,Witten-0312171}.
Similarly, the sub-leading soft divergence for Yang-Mills amplitudes was obtained by same analysis in \cite{Casali-1404.5551}.
As pointed out in~\cite{Bianchi-He-Huang-Wen-1406.5155}, the sub-leading divergence is vanishing in pure Yang-Mills amplitudes.
However, we will see that the sub-leading Yang-Mills soft operators are necessary for ``KLT-construction'' of gravity soft operators in this paper.
On the other hand, a lot of remarkable progress has also been made on the understanding of the soft theorem from the viewpoint of symmetry principles.
The leading and the sub-leading soft graviton theorems are understood as Ward identities of BMS symmetry \cite{Strominger-1312.2229,He-Lysov-Mitra-Strominger-1401.7026,Kapec-Lysov-Pasterski-Strominger-1406.3312}.
The leading \cite{He-Mitra-Porfyriadis-Strominger-1407.3789} and sub-leading \cite{Lysov-Pasterski-Strominger-1407.3814} soft photon theorems were interpreted as asymptotic symmetries of $\cal S$-matrix in massless QED.

Now aspects of soft theorems have been investigated along many different directions \cite{Casali-1404.5551,Cachazo-Yuan-1405.3413,
He-Huang-Wen-1405.1410,
Bianchi-He-Huang-Wen-1406.5155,
Bern-Davies-Nohle-1405.1015,
Bern-Davies-Vecchia-Nohle-1406.6987,
Broedel-Leeuw-Plefka-Rosso-1406.6574,
White-1406.7184,Schwab-Volovich-1404.7749,
Afkhami-Jeddi-1405.3533,
Zlotnikov-1407.5936,Kalousios-Rojas-1407.5982,
Schwab-1406.4172,Geyer-Lipstein-Mason-1406.1462,
Larkoski-1405.2346,
Lysov-Pasterski-Strominger-1407.3814,
White-1103.2981,
Adamo-Casali-Skinner-1405.5122,
Strominger-1312.2229,
He-Lysov-Mitra-Strominger-1401.7026,
Kapec-Lysov-Pasterski-Strominger-1406.3312,
He-Mitra-Porfyriadis-Strominger-1407.3789,
Naculich-1407.7836,
Du-Feng-Fu-Wang-1408.4179,
Laenen-Stavenga-White-0811.2067,Laenen-Magnea-Gerben Stavenga-White-1010.1860,Oxburgh-White-1210.1110}.
Loop corrections of soft theorems were investigated in both gravity and Yang-Mills \cite{Cachazo-Yuan-1405.3413,He-Huang-Wen-1405.1410,Bianchi-He-Huang-Wen-1406.5155,Bern-Davies-Nohle-1405.1015}.
Various different methods were used to derive the soft theorems, including Feynman diagram approach \cite{White-1406.7184} and conformal symmetry approach \cite{Larkoski-1405.2346} for Yang-Mills, the Poincar\'e symmetry and gauge symmetry approach \cite{Broedel-Leeuw-Plefka-Rosso-1406.6574,Bern-Davies-Vecchia-Nohle-1406.6987}, ambitwistor string approach \cite{Geyer-Lipstein-Mason-1406.1462} for Yang-Mills and gravity.
The new soft graviton theorem was also obtained from the soft gluon theorem  via the Kawai-Lewellen-Tye (KLT) relations \cite{Du-Feng-Fu-Wang-1408.4179}.
The scattering equations \cite{Cachazo-He-Yuan-1306.2962,Cachazo-He-Yuan-1306.6575,Cachazo-He-Yuan-1307.2199,Cachazo-He-Yuan-1309.0885}, or Cachazo-He-Yuan (CHY) formulae of tree-level S-matrices of gluons as well as gravitons, were used to study the soft theorems in arbitrary number of dimensions.
The soft divergence in string theory was also investigated in \cite{Bianchi-He-Huang-Wen-1406.5155,Schwab-1406.4172}.

It is natural and interesting to investigate soft divergences of amplitudes in supersymmetric theories.
In this paper, we mainly focus on the soft theorems in the maximally supersymmetric theories, i.e. ${\cal N}=4$ super-Yang-Mills (SYM) theory and ${\cal N}=8$ supergravity (SUGRA), in $\it 4$ spacetime dimensions.
Great progress on analytical calculations of the scattering amplitudes, in particular at tree-level, has been achieved in ${\cal N}=4$ SYM and ${\cal N}=8$ SUGRA.
A lot of remarkable and interesting structures of scattering amplitudes were discovered \cite{Drummond-Henn-Plefka-0902.2987,Drummond-Henn-Korchemsky-Sokatchev-0807.1095,
Brandhuber-Heslop-Travaglini-0807.4097,
Korchemsky-Sokatchev-0906.1737,Korchemsky-Sokatchev-1002.4625,
Beisert-1012.3982}.
Many novel methods for perturbative scattering amplitudes were proposed, and a large number of amplitudes were also analytically computed in various theories \cite{Elvang-Huang-1308.1697,
Henn-Plefka-Book,Nima-Cachazo-Kaplan-0808.1446,Dixon-Review-1105.0771,
Bianchi-Elvang-Freedman-0805.0757,
Elvang-Freedman-Kiermaier-0911.3169,
Drummond-Spradlin-Volovich-Wen-0901.2363,
Dixon-Henn-Plefka-Schuster-1010.3991,
Nandan-Wen-1204.4841,
Boucher-Veronneau-Larkoski-1108.5385,
Britto-Feng-Roiban-Spradlin-Volovich-0503198,
Drummond-Henn-0808.2475,
Kallosh-Lee-Rube-0811.3417,
Bjerrum_Bohr-Engelund-1002.2279,Grisaru-Pendleton-NPB1977,Grisaru-Pendleton-Nieuwenhuizen-PRD1977}.
For example, by solving super-BCFW recursion relation \cite{Britto-Cachazo-Feng-0412308,Britto-Cachazo-Feng-Witten-0501052,Nima-Kaplan-0801.2385,Nima-Cachazo-Kaplan-0808.1446},
the compact analytical formulae for all tree amplitudes were presented in ${\cal N}=4$ SYM \cite{Drummond-Henn-0808.2475}, also in massless QCD with up to four quark-anti-quark pairs \cite{Dixon-Henn-Plefka-Schuster-1010.3991}.
Scattering amplitudes in massless QCD were also studied in \cite{Melia-1312.0599,Melia-1304.7809}.
Scattering amplitudes in massless QCD were also studied in \cite{Melia-1312.0599,Melia-1304.7809}.
These previous works provide a solid foundation for our study.

We will systematically study the soft theorems in ${\cal N}=4$ SYM and ${\cal N}=8$ SUGRA using super-BCFW recursion relations in this work.
We will pay special attention to soft gravitino divergence, soft gravi-photon divergence for SUGRA and soft gluino divergence for SYM.




This paper is organized as follows.
In the next section, we briefly review soft theorems in both gravity and Yang-Mills.
In section \ref{Sec-N=4}, we present the super soft theorem in ${\cal N}=4$ super-Yang-Mills with a rigorous proof in detail.
There we also provide some lower-point amplitudes to test the validity of the super soft theorem, especially for soft gluino divergence.
In section \ref{Sec-N=8}, we study the super soft theorem in ${\cal N}=8$ supergravity in detail.
As examples, we show that the super soft theorem is consistent with SUSY Ward identity in the MHV sector.
Several four-boson amplitudes are also examined exactly.
In section \ref{Sec-Conclusion}, we conclude this paper with some brief discussions.
Appendix \ref{Sec-N=4-alternative} provides an alternative derivation of soft theorem in ${\cal N}=4$ SYM.
Appendix \ref{Sec-N=8-NNLO-Soft-Operators} gives calculational details of the sub-sub-leading soft operator in ${\cal N}=8$ SUGRA.

\section{New soft graviton theorem and soft gluon theorem}\label{Sec-Review}
In this section we briefly review the new soft graviton theorem \cite{Cachazo-Strominger-1404.4091} and the soft gluon theorem \cite{Casali-1404.5551,Berends-Giele-1989}.
We also show that each order soft operator in gravity may be expressed as a double copy of Yang-Mills soft operators with gauge freedom.
\subsection{Cachazo-Strominger's new soft graviton theorem}
An on-shell $(n+1)$-point scattering amplitude including an external graviton with
momentum $k_s$ may be denoted
\begin{align}\label{}
  {\cal M}_{n+1} \,=\,  {\cal M}_{n+1}(k_1,\ldots,k_n,k_s).
\end{align}
In the soft limit $k_s\to 0$, the amplitude ${\cal M}_{n+1}$ behaves as
\begin{align}\label{new-soft-gravitom-th-1}
  {\cal M}_{n+1}(k_1,\ldots,k_n,k_s) \,=\,
  \Big(S^{(0)} + S^{(1)} + S^{(2)} \Big) {\cal M}_{n}(k_1,\ldots,k_n) + {\cal O}(k_s^2).
\end{align}
The soft operators are given by
\begin{align}\label{}
   S^{(0)} \,&=\, \sum_{a=1}^{n}{E_{\mu\nu} k_a^\mu k_a^\nu \over k_s\cdot k_a},
   \\
   S^{(1)} \,&=\, -i\sum_{a=1}^{n}
   {E_{\mu\nu} k_a^\mu \big(k_{s\sigma}J_a^{\sigma\nu}\big) \over k_s\cdot k_a},
   \\
   S^{(2)} \,&=\, -{1\over2}\sum_{a=1}^{n} {E_{\mu\nu}
   \big(k_{s\rho}J_a^{\rho\mu}\big) \big(k_{s\sigma}J_a^{\sigma\nu}\big) \over k_s\cdot k_a},
\end{align}
where $E_{\mu\nu}$ is the polarization tensor of the soft graviton $s$ and $J_a^{\mu\nu}$ is the total angular momentum of the $a$th external leg.
It is easy to check that all the soft operators are gauge invariant \cite{Cachazo-Strominger-1404.4091}.
The leading soft factor $S^{(0)}$ proposed by Steven Weinberg is uncorrected by all loop orders \cite{Weinberg1965} while sub-leading and sub-sub-leading soft operators are not, as discussed in \cite{Cachazo-Yuan-1405.3413,He-Huang-Wen-1405.1410,Bianchi-He-Huang-Wen-1406.5155,Bern-Davies-Nohle-1405.1015,Bern-Davies-Vecchia-Nohle-1406.6987}.

In the spinor-helicity formulism, the momentum vector $k^\mu$ of an on-shell massless particle may be represented as a bispinor, i.e.,
\begin{align}\label{}
  k_{\alpha\dot\alpha} \,=\, k_\mu\sigma^\mu_{\alpha\dot\alpha} \,=\, \lambda_\alpha\tilde\lambda_{\dot\alpha}.
\end{align}
Introducing an infinitesimal soft parameter $\epsilon$, one can write soft limit of the momentum $k_s$ of soft particle as
\begin{align}\label{}
  k_s \,\to\, \epsilon\, k_s = \epsilon\, \lambda_s\tilde\lambda_s
  \qquad\Longrightarrow\qquad
  \lambda_s \to \epsilon^{\delta}\lambda_s,\quad \tilde\lambda_s\to\epsilon^{1-\delta}\tilde\lambda_s.
\end{align}
Here different choices of $\delta$ in physical amplitudes can be linked to each other via the little group transformation, i.e.,
\begin{align}\label{}
  {\cal M}\big(\{t\lambda_s, t^{-1}\tilde\lambda_s, h_s\}\big) \,&=\,
  t^{-2h_s} {\cal M}\big(\{\lambda_s, \tilde\lambda_s, h_s\}\big),
\end{align}
where $h_s$ is helicity of the particle $s$.
In this paper, one employs the {\it holomorphic soft limit} \cite{Nima-Cachazo-Kaplan-0808.1446}:
\begin{align}\label{holomorphic-soft-limit-1}
  \lambda_s \to \epsilon\,\lambda_s
\end{align}
in which only the holomorphic spinor $\lambda_s$ tends to zero while the anti-holomorphic spinor $\tilde\lambda_s$ remains unchangeable.
The new soft graviton theorem \eqref{new-soft-gravitom-th-1} is then
\begin{align}\label{}
  {\cal M}_{n+1}(\ldots,\{\epsilon\lambda_s, \tilde\lambda_s\}) \,=\,
  \bigg({1\over\epsilon^3}S^{(0)} + {1\over\epsilon^2}S^{(1)} + {1\over\epsilon}S^{(2)} \bigg)
  {\cal M}_{n} + {\cal O}(\epsilon^0).
\end{align}
In the spinor-helicity formulism\footnote{In this paper, we mainly follow the notation of ref.~\cite{Cachazo-Strominger-1404.4091}.
The spinor products are defined as
$\displaystyle
\braket{i,j}=\epsilon^{\alpha\beta} \lambda_{i\alpha} \lambda_{j\beta}
=\lambda_{i\alpha} \lambda_j^\alpha
$ and $\displaystyle
  [i,j]=\epsilon^{\dot\alpha\dot\beta} \tilde\lambda_{i\dot\alpha} \tilde\lambda_{j\dot\beta}
  = \tilde\lambda_{i\dot\alpha} \tilde\lambda_j^{\dot\alpha}
$,
and we use the convention $s_{ij}=\braket{i,j}[i,j]$ whic is different from QCD convention.},
the soft operators are given by
\begin{align}\label{}
  S^{(0)} \,&=\, \sum_{a=1}^{n}
  {[s,a] \braket{x,a}\braket{y,a} \over \braket{s,a} \braket{x,s} \braket{y,s}},
  \\
  S^{(1)} \,&=\, {1\over 2}\sum_{a=1}^{n} {[s,a] \over \braket{s,a}}
  \left( {\braket{x,a} \over \braket{x,s}} + {\braket{y,a} \over \braket{y,s}} \right)
  \tilde\lambda_{s\dot\alpha} {\partial\over\partial\tilde\lambda_{a\dot\alpha}},
  \\
  S^{(2)} \,&=\,{1 \over 2} \sum_{a=1}^{n} {[s,a]  \over \braket{s,a} }
  \tilde\lambda_{s\dot\alpha} \tilde\lambda_{s\dot\beta}
  {\partial^2 \over \partial\tilde\lambda_{a\dot\alpha} \partial\tilde\lambda_{a\dot\beta}}.
\end{align}
Here one has assigned soft graviton the helicity $h_s=+2$, just a convention.
Spinors $\lambda_x$, $\lambda_y$ are two arbitrary choosen reference spinors and the freedom in this choice is equivalent to the gauge freedom.

\subsection{Soft gluon theorem in Yang-Mills theory}
The similar soft behavior of the scattering amplitudes appears also in Yang-Mills theory \cite{Casali-1404.5551,Berends-Giele-1989}.
In the soft limit of the momentum of a gluon, $k_s\to 0$, an on-shell {\it color-ordered} Yang-Mills amplitude ${\cal A}_{n+1}$ becomes
\begin{align}\label{soft-gluon-theorem-1}
  {\cal A}_{n+1}(k_1,\ldots,k_n,k_s) \,=\,
  \Big( S_{\rm YM}^{(0)} + S_{\rm YM}^{(1)} \Big)
  {\cal A}_{n}(k_1,\ldots,k_n)
  + {\cal O}(k_s),
\end{align}
where the leading soft (eikonal) factor \cite{Berends-Giele-1989} is
\begin{align}\label{}
  S_{\rm YM}^{(0)} \,\equiv\, \sum_{a=1,n} {E_\mu k_a^\mu \over k_s\cdot k_a},
\end{align}
while the sub-leading soft operator is given by
\begin{align}\label{}
  S_{\rm YM}^{(1)} \,\equiv\, -i\sum_{a=1,n} {E_\mu k_{s\nu} J_a^{\mu\nu} \over k_s\cdot k_a},
\end{align}
with $E_\mu$ the polarization vector of soft gluon.

In the spinor-helicity formulism, employing the {\it holomorphic soft limit} \eqref{holomorphic-soft-limit-1}, the soft gluon theorem \eqref{soft-gluon-theorem-1} may be expressed as
\begin{align}\label{soft-gluon-theorem-2}
  {\cal A}_{n+1}(\ldots,\{\epsilon\lambda_s, \tilde\lambda_s\})
  \,&=\, \left( {1\over\epsilon^2}S^{(0)} + {1\over\epsilon}S^{(1)} \right)
  {\cal A}_n + {\cal O}(\epsilon^0).
\end{align}
Taking the helicity of the soft gluon $h_s=+1$ as a convention, the soft operators may be written as
\begin{align}\label{}
  S_{\rm YM}^{(0)} \,&=\,
  {\braket{x,n} \over \braket{s,n}\braket{x,s}} + {\braket{x,1} \over \braket{s,1}\braket{x,s}},
  \\
  S_{\rm YM}^{(1)} \,&=\, {1 \over \braket{n,s} }
  \tilde\lambda_{s\dot\alpha}\,{\partial\over\partial\tilde\lambda_{n\dot\alpha} }
  \,+\, {1 \over \braket{s,1} }
  \tilde\lambda_{s\dot\alpha}\,{\partial\over\partial\tilde\lambda_{1\dot\alpha} },
\end{align}
with $\lambda_x$, $\lambda_y$ arbitrary choosen reference spinors and the freedom in this choice is equivalent to the gauge freedom.

It is important to note that two amplitudes in the soft theorem are both {\it unstripped}.
In other words, amplitudes ${\cal A}_{n+1}$ and ${\cal A}_{n}$ in eq.~\eqref{soft-gluon-theorem-2} contain respective momentum conservation delta functions.
With this in mind, we can remove dependence of anti-holomorphic spinors $\tilde\lambda_1$ and $\tilde\lambda_n$ in these two amplitudes by imposing momentum conservation delta functions appropriately.
This implies that the sub-leading soft divergence vanishes in color-ordered Yang-Mills amplitudes \cite{Bianchi-He-Huang-Wen-1406.5155}.
As we will see immediately, however, that the operator $\displaystyle S_{\rm YM}^{(1)}$ is necessary for constructing gravity soft operators from Yang-Mills soft operators.

\subsection{Gravity soft operators as double copy of Yang-Mills soft operators}\label{Sec-soft-operator-double-copy-1}
There exists a remarkable relation between gravity amplitudes and Yang-Mills amplitudes.
At tree level, Kawai, Lewellen and Tye found that one can express a closed string amplitude as a sum of the square of open string amplitudes \cite{Kawai-Lewellen-Tye-1985}.
In field theory limit, this relation expresses a gravity amplitude as a sum of the square of color-ordered Yang-Mills amplitudes.
The similar relation also exists between gravity soft operators and Yang-Mills soft operators.
In \cite{He-Huang-Wen-1405.1410}, the gravity soft operators were expressed as a double copy of Yang-Mills soft operators with a special gauge choice which associated with the special choice of shifted external legs in BCFW recursion.
In this subsection, we rewrite this relation with {\it gauge freedom}.

First of all, for the sake of convenience, introduce two notations\footnote{Here the first notation ${\frak S}^0$ is just famous eikonal factor in Yang-Mills amplitudes \cite{Dixon-Review-96,Bern-Grant-9904026,Bern-2002-Rev-0206071} and the `$x$' denotes the reference spinor $\lambda_x$ in spinor representation $\displaystyle E^+_{\alpha\dot\alpha}(\lambda_s,\tilde\lambda_s,\lambda_x)$ of the polarization vector $E^+_\mu$ \cite{Witten-0312171}.}
\begin{align}
  \label{soft-operator-seed-0}
  {\frak S}^0 (x,s,a) \,&\equiv\, {E_\mu^+(\lambda_x) k_a^\mu \over k_s\cdot k_a}
  \,=\, {\braket{x,a} \over \braket{x,s}\braket{s,a}},  \\
  \label{soft-operator-seed-1}
  {\frak S}^1 (s,a) \,&\equiv\, -i{E_\mu^+(\lambda_x) k_{s\nu} J_a^{\mu\nu} \over k_s\cdot k_a}
  \,=\, {1 \over \braket{s,a}} \tilde\lambda_{s\dot\alpha}{\partial\over\partial\tilde\lambda_{a\dot\alpha}},
\end{align}
which are the fundamental building blocks for constructing gravity soft operators.
Employing these notations, one can write the Yang-Mills soft operators as
\begin{align}\label{}
  S_{\rm YM}^{(0)} \,&=\, {\frak S}^0(x,s,n) + {\frak S}^0(x,s,1),
  \\
  S_{\rm YM}^{(1)}(1,s,n) \,&=\, {\frak S}^1(s,1) - {\frak S}^1(s,n).
\end{align}

Let us note a simple relation that expresses a graviton polarization tensor as the product of gluon polarization vectors with same momentum, i.e.,
\begin{align}\label{}
  E_{\mu\nu}^\pm(k) \,&=\, E_\mu^\pm(k) \times E_{\nu}^\pm(k) + E_\nu^\pm(k) \times E_{\mu}^\pm(k).
\end{align}
Here $E_{\mu\nu}$ have been written in a symmetric form.
By making use of this relation, the leading soft operator in gravity can be written as:
\begin{align}\label{}
   S^{(0)} \,&=\, \sum_{a=1}^{n}{2\big(E_{\mu}^+(\lambda_x) E_{\nu}^+(\lambda_y) \big) k_a^\mu k_a^\nu \over k_s\cdot k_a}
   \nonumber\\
   &=\, \sum_{a=1}^{n} \big(2k_s\cdot k_a\big)
   \bigg( {E_{\mu}^+(\lambda_x) k_a^\mu\over k_s\cdot k_a} \bigg)
   \bigg( {E_{\nu}^+(\lambda_y) k_a^\nu \over k_s\cdot k_a} \bigg)
   \nonumber\\
   &=\, \sum_{a=1}^{n} s_{sa}{\frak S}^0(x,s,a) {\frak S}^0(y,s,a)
   \label{leading-soft-gravity-KLT-1}
\end{align}
where $s_{sa} = 2k_s\cdot k_a = \braket{s,a}[s,a]$ and the $\lambda_x$ and $\lambda_y$ are arbitrary reference spinors and the freedom in this choice is equivalent to the gauge freedom.
This relation was presented in \cite{Berends-Giele-Kuijf-PLB-1988,Bern-Grant-9904026,Bern-2002-Rev-0206071} and derived in \cite{Bjerrum-Bohr-Damgaard-Sondergaard-Vanhove-1010.3933,Du-Feng-Fu-Wang-1408.4179} by KLT realtion \cite{Kawai-Lewellen-Tye-1985,Bjerrum-Bohr-Damgaard-Feng-Sondergaard-1005.4367,Bjerrum-Bohr-Damgaard-Feng-Sondergaard-1007.3111}.

The sub-leading soft graviton operator can be expressed as
\begin{align}\label{}
   S^{(1)} \,&=\, -i\sum_{a=1}^{n} (k_s\cdot k_a) \left(
   {E_{\mu}^+(\lambda_x) k_a^\mu \over k_s\cdot k_a}
   {E_{\nu}^+(\lambda_y) k_{s\sigma} J_a^{\sigma\nu} \over k_s\cdot k_a}
   \,+\,
   (x \leftrightarrow y)
   \right)
   \nonumber\\
   &=\, \frac12 \sum_{a=1}^{n} s_{sa}
   \Big( {\frak S}^0(x,s,a) + {\frak S}^0(y,s,a) \Big){\frak S}^1(s,a)
\end{align}
This relation with a special gauge choice was derived in \cite{Du-Feng-Fu-Wang-1408.4179} by KLT realtion \cite{Kawai-Lewellen-Tye-1985,Bjerrum-Bohr-Damgaard-Feng-Sondergaard-1005.4367,Bjerrum-Bohr-Damgaard-Feng-Sondergaard-1007.3111}.

Similarly, the sub-sub-leading soft operator may be written as
\begin{align}\label{}
   S^{(2)}
   \,&=\, {1\over2}\sum_{a=1}^{n} s_{sa}\, {\frak S}^1(s,a) {\frak S}^1(s,a)
\end{align}
It is important to notice that here the operator product ${\frak S}^1(s,a) {\frak S}^1(s,a)$ should be understood as:
\begin{align}\label{}
  {\frak S}^1 (s,a) {\frak S}^1 (s,a) \,&=\, {1 \over \braket{a,s}^2} \tilde\lambda_{s\dot\alpha}\tilde\lambda_{s\dot\beta}
  {\partial^2\over\partial\tilde\lambda_{a\dot\alpha} \partial\tilde\lambda_{a\dot\beta}}.
\end{align}
In another words, the differential only acts on the amplitudes.

\section{Soft theorem in ${\cal N}=4$ super-Yang-Mills theory}\label{Sec-N=4}
We turn to study the soft theorems in supersymmetric theories.
In this section, we present the soft theorem in ${\cal N}=4$ super-Yang-Mills theory with a rigorous proof at tree level.
We also give some lower-point amplitudes examples to demonstrate the validity of super soft theorem, in particular the soft gluino divergence.

Let us begin with a very brief introduction of ${\cal N}=4$ SYM and the superamplitudes in on-shell superspace.
The ${\cal N}=4$ on-shell field consists of 8 bosons and 8 fermions and one can write it out as
\begin{align}\label{}
  h={1}:\quad & 1~\text{gluon}~g^+  \nonumber\\
  h={1\over2}:\quad & 4~\text{gluinos}~\Gamma_A  \nonumber\\
  h=0:\quad & 6~\text{scalars}~S_{AB}  \\
  h=-{1\over2}:\quad & 4~\text{gluinos}~\bar\Gamma^A  \nonumber\\
  h=-1:\quad & 1~\text{gluon}~g^- \nonumber
\end{align}
Here $A, B, \ldots = 1,2,3,4$ are ${\rm SU}(4)$ R-symmetry indices and the scalar $S_{AB}$ is antisymmetric in indices $A$, $B$.
The ${\cal N}=4$ on-shell superfield can be expanded as follows \cite{Nair-1988}:
\begin{align}\label{}
  \Phi(p,\eta) \,=\, g^+(p) + \eta^A \Gamma_A(p) + {1\over 2!} \eta^{A}\eta^{B}S_{AB}(p)
  + {1\over 3!} \eta^{A}\eta^{B}\eta^{C} \epsilon_{ABCD} \bar\Gamma^{D}(p)
  + \eta^{1}\eta^{2}\eta^{3}\eta^{4}\, g^-(p).
\end{align}
Here Grassmann odd variables $\eta^A$ transforms in a fundamental representation of the ${\rm SU}(4)$ R-symmetry.

In super-momentum space, a {\it color-ordered superamplitude} is a function of spinors $\lambda_a,\,\tilde\lambda_a$ (or momentum $p_a$) and Grassmann variables $\eta_a$, i.e.,
\begin{align}\label{}
  {\mathscr A}_{n} \,&\equiv\, {\mathscr A}_n ( \Phi_1, \ldots, \Phi_n )
  \,=\,
  {\mathscr A}_{n}
  \big( \{\lambda_1,\tilde\lambda_1,\eta_1\}, \ldots, \{\lambda_n,\tilde\lambda_n,\eta_n\}\big).
\end{align}
The component field amplitudes are then obtained by projecting upon the relevant terms in the $\eta_i$ expansion of the superamplitude. For detail, see \cite{Elvang-Huang-1308.1697,Henn-Plefka-Book,Nima-Cachazo-Kaplan-0808.1446}.

\subsection{Super soft theorem in ${\cal N}=4$ SYM}
Here we derive the soft theorem in ${\cal N}=4$ SYM with the help of super-BCFW recursion relation \cite{Britto-Cachazo-Feng-0412308,Britto-Cachazo-Feng-Witten-0501052,Nima-Cachazo-Kaplan-0808.1446} in the spinor-helicity formulism \cite{Dixon-Review-96,Dixon-Review-2013,Xu-Zhang-Chang-NPB-1987}.
Let us choose the soft particle and its adjacent particle to sfift:
\begin{align}\label{super-shift-1}
  \lambda_s(z)=\lambda_s+z\lambda_n,\quad \tilde\lambda_n(z)=\tilde\lambda_n-z\tilde\lambda_s,\quad \eta_n(z)=\eta_n - z\eta_s.
\end{align}
These shifts preserve the total momentum and super-momentum.
Super-BCFW recursion gives:
\begin{align}\label{BCFW-N=4-1}
  {\mathbb A}_{n+1} \,=\, \sum_{a=1}^{n-2} \int d^4\eta_I\, &
  {\mathbb A}_L\big( \{\lambda_s(z^*),\tilde\lambda_s, \eta_s\},
  1, \ldots, a, \{I(z^\ast), \eta_I\} \big)
  \nonumber\\
  &\times \frac{1}{P_I^2}
  {\mathbb A}_R\big( \{-I(z^\ast), \eta_I\},a+1, \ldots, n-1, \{\lambda_n,\tilde\lambda_n(z^\ast), \eta_n(z^*)\} \big).
\end{align}
Here the integral over $\eta_I$ denotes the sum over intermediate states in ordinary BCFW recursion \cite{Nima-Cachazo-Kaplan-0808.1446}.
The blackboard-bold style denotes the stripped superamplitude,
\begin{align}\label{}
  {\mathscr A}_n \,&=\, {\mathbb A}_n\, \delta^4(p).
\end{align}
According to different Grassmann odd degrees, one can decompose the superamplitude into various N$^k$MHV sectors, i.e.,
\begin{align}\label{BCFW-N=4-3}
  {\mathbb A}^{\rm N^{\it k}MHV}_{n+1} \,=\,&
  \int d^4\eta_P\, {\mathbb A}_3^{\overline{\rm MHV}}(z^*) \frac{1}{P^2}
  {\mathbb A}_{n}^{\rm N^{\it k}MHV} (z^*)
  \nonumber\\
  &+ \sum_{m=0}^{k-1}\sum_{a=4}^n
  \int d^4\eta_{P_a} {\mathbb A}_a^{\rm N^{\it m}MHV}(z_a) \frac{1}{P_a^2}
  {\mathbb A}_{n-a+3}^{{\rm N}^{(k-m-1)}{\rm MHV}} (z_a).
\end{align}
As shown explicitly in \cite{Cachazo-Strominger-1404.4091}, the singular terms only come from the term with $a=1$ in eq.~\eqref{BCFW-N=4-1}, or the first term of the right hand side in eq.~\eqref{BCFW-N=4-3} under the holomorphic soft limit \eqref{holomorphic-soft-limit-1}.
So we drop the terms from contributions with $a>1$, and write
\begin{align}\label{BCFW-N=4-singular-2}
  {\mathbb A}_{n+1}  \,=\, \int d^4\eta_I\, &
  {\mathbb A}_3^{\overline{\rm MHV}} \big( \{\hat s(z^\ast), \eta_s\},
  \{1,\eta_1\}, \{I(z^\ast), \eta_I(z^\ast)\} \big)
  \nonumber\\
  & \times \frac{1}{P_I^2}
  {\mathbb A}_n \big( \{-I(z^\ast), \eta_I(z^\ast)\}, \{2,\eta_2\}, \ldots, \{\hat n(z^\ast), \eta_n(z^\ast)\} \big).
\end{align}
Graphically,
\begin{center}
\begin{tikzpicture}[scale=1]
  \draw[line width=1.7pt] (0,0) circle (20pt);
  \fill (0,0) circle (0pt)  node {\large${\mathbb A}_{n+1}$}; 

  \draw[color=red,line width=0.9pt](0:20pt) -- (0:55pt) node[right] { $\hat n$};
  \draw[color=red,line width=0.9pt](180:20pt)--(180:55pt) node[left] { $\hat s$};

  \draw[line width=0.6pt](140:20pt) -- (140:60pt) node[above] {$1$};
  \draw[line width=0.6pt](40:20pt) -- (40:60pt) node[above] {$n\!-\!1$};

  \fill[color=blue] (90:36pt)circle (1.0pt);
  \fill[color=blue] (110:36pt)circle (1.0pt);
  \fill[color=blue] (70:36pt)circle (1.0pt);

  \coordinate [label=0:\parbox{35mm}{ {\Large$\displaystyle =\int d^4\eta_I$} }] (sum) at (70pt,0pt);

  \draw[xshift=210pt,line width=1.7pt] (0,0) circle (18pt);
  \fill[xshift=210pt] (0,0) circle (0pt)  node {\Large${\mathbb A}_L$}; 

  \draw[xshift=210pt,color=red,line width=0.9pt](180:18pt)--(180:52pt) node[left] { $\hat s$};

  \draw[xshift=210pt,line width=0.9pt](90:18pt) -- (90:47pt) node[right] {$1$};

  \draw[xshift=210pt,color=blue,line width=0.7pt](0:18pt) -- (0:70pt);

  \draw[xshift=320pt,line width=1.7pt] (0,0) circle (18pt);
  \fill[xshift=320pt] (0,0) circle (0pt)  node {\Large${\mathbb A}_R$}; 

  \draw[xshift=320pt,color=blue,line width=0.7pt](180:18pt) -- (180:80pt);

  \draw[xshift=320pt,color=red,line width=0.9pt](0:18pt) -- (0:52pt) node[right] {$\hat n$};

  \draw[xshift=320pt,line width=0.6pt](140:18pt) -- (140:60pt) node[above] {$2$};
  \draw[xshift=320pt,line width=0.6pt](40:18pt) -- (40:60pt) node[above] {$n \!-\! 1$};
  \fill[xshift=320pt,color=blue] (90:32pt)circle (1.0pt);
  \fill[xshift=320pt,color=blue] (110:32pt)circle (1.0pt);
  \fill[xshift=320pt,color=blue] (70:32pt)circle (1.0pt);

  \fill[xshift=210pt,color=red] (14:35pt) circle (0pt)  node {$\eta_I$}; 
  \fill[xshift=330pt,color=red] (169:45pt) circle (0pt)  node {$\eta_I$}; 

\end{tikzpicture}
\end{center}
Here we have
\begin{align}\label{}
  P_I^2 \,&=\, \big(k_s + k_1\big)^2 \,=\, \braket{s,1}[s,1],  \\
  z^\ast \,&=\, -{\braket{s,1} \over \braket{n,1}},  \\
  \lambda_I \,&=\, \lambda_1, \\
  \tilde\lambda_I \,&=\, \tilde\lambda_1 + {\braket{n,s} \over \braket{n,1}}\tilde\lambda_s.
\end{align}
In on-shell resursions of tree-level amplitudes, three-point amplitudes are seeds for generating higher-point amplitudes.
In on-shell superspace, the three-point superamplitudes of ${\cal N}=4$ SYM are given by
\begin{align}
  \label{N=4-3pt-MHV-1}
  {\mathscr A}_3^{\rm MHV} \,&=\, {\delta^4(p) \over \braket{12}\braket{23}\braket{31}}
  \delta^8\Big(\sum\nolimits_{a=1}^3\lambda_a^\alpha\eta_a^A\Big),
  \\
  \label{N=4-3pt-anti-MHV-1}
  {\mathscr A}_3^{\overline{\rm MHV}} \,&=\, {\delta^4(p) \over [12][23][31] }
  \delta^4 \big([12] \eta_3^A + [23] \eta_1^A + [31] \eta_2^A \big).
\end{align}
Then it is easy to get the left three-point superamplitude in eq. \eqref{BCFW-N=4-singular-2}
\begin{align}\label{N=4-BCFW-3pt-1}
  {\mathbb A}_3^{\overline{\rm MHV}}
  \big( \{\lambda_s(z^*),\tilde\lambda_s, \eta_s\}, \{1,\eta_1\}, \{I, \eta_I\} \big)
  \,&=\, {\braket{n,1}\, [s,1] \over \braket{n,s} }
  \delta^4 \Big( \eta_I - {\braket{n,s} \over \braket{n,1}} \eta_s - \eta_1 \Big).
\end{align}
Notice that the left superamplitude in eq.~\eqref{BCFW-N=4-singular-2} is three-point anti-MHV. In fact, this corresponds to the super-shift \eqref{super-shift-1} and in this case the helicity of soft gluon takes positive one \cite{Nandan-Wen-1204.4841}.
Inserting the 3-point superamplitude \eqref{N=4-BCFW-3pt-1} into eq.~\eqref{BCFW-N=4-singular-2} and computing the integral over $\eta_I$ give
\begin{align}\label{}
  {\mathbb A}_{n+1} &= {\braket{n,1} \over \braket{n,s}\braket{s,1} }\,
  {\mathbb A}_n\Big( \{\lambda_1, \tilde\lambda_1 + {\braket{n,s} \over \braket{n,1}}\tilde\lambda_s, \eta_1 + {\braket{n,s} \over \braket{n,1}}\eta_s\},
  \{\lambda_2, \tilde\lambda_2, \eta_2\}, \ldots,
  \{\lambda_n,\tilde\lambda_n + {\braket{s,1} \over \braket{n,1}}\tilde\lambda_s,
  \eta_n + {\braket{s,1} \over \braket{n,1}}\eta_s\} \Big).
\end{align}
Dressing both sides of the above equation in respective appropriate momentum conservation delta functions, one obtains
\begin{align}\label{}
  {\mathscr A}&_{n+1} \,=\, {\braket{n,1} \over \braket{n,s}\braket{s,1} }
  \nonumber\\
  &\times
  {\mathscr A}_n\Big( \{\lambda_1, \tilde\lambda_1 + {\braket{n,s} \over \braket{n,1}}\tilde\lambda_s, \eta_1 + {\braket{n,s} \over \braket{n,1}}\eta_s\},
  \{\lambda_2, \tilde\lambda_2, \eta_2\},
  \ldots,
  \{\lambda_n,\tilde\lambda_n + {\braket{s,1} \over \braket{n,1}}\tilde\lambda_s,
  \eta_n + {\braket{s,1} \over \braket{n,1}}\eta_s\} \Big).
\end{align}
In the holomorphic soft limit \eqref{holomorphic-soft-limit-1},
\begin{align}\label{}
  {\mathscr A}&_{n+1}(\epsilon) \,=\, {1\over\epsilon^2} {\braket{n,1} \over \braket{n,s}\braket{s,1} }
  \nonumber\\
  \times &
  {\mathscr A}_n\Big( \{\lambda_1,
  \tilde\lambda_1 + \epsilon {\braket{n,s} \over \braket{n,1}}\tilde\lambda_s,
  \eta_1 + \epsilon {\braket{n,s} \over \braket{n,1}}\eta_s\},
  \{\lambda_2, \tilde\lambda_2, \eta_2\},
  \ldots,
  \{\lambda_n,\tilde\lambda_n + \epsilon {\braket{s,1} \over \braket{n,1}}\tilde\lambda_s,
  \eta_n + \epsilon {\braket{s,1} \over \braket{n,1}}\eta_s\} \Big).
\end{align}
Performing Taylor expansion at $\epsilon=0$, we obtain the soft theorem\footnote{
The same result was obtained in the Grassmannian formulation in \cite{Rao-1410.5047}.
}
\begin{align}\label{soft-theorem-N=4-1}
  {\mathscr A}_{n+1}(\epsilon) \,&=\,
  \left({1\over\epsilon^2} {\cal S}_{\rm SYM}^{(0)} + {1\over\epsilon} {\cal S}_{\rm SYM}^{(1)}\right)
  {\mathscr A}_n
  \,+\, {\cal O}(\epsilon^0)
\end{align}
where
\begin{align}\label{}
  {\cal S}_{\rm SYM}^{(0)} \,&=\, {\braket{n,1} \over \braket{n,s}\braket{s,1}} \,=\, S_{\rm YM}^{(0)},
  \\
  {\cal S}_{\rm SYM}^{(1)} \,&=\, S_{\rm YM}^{(1)}
  \,+\, \eta_s^A {\cal F}^{(0)}_A,
  \qquad
  {\cal F}^{(0)}_A \,\equiv\,
  {1 \over \braket{s,1}} {\partial\over\partial\eta_1^A}
  +  {1 \over \braket{n,s}}  {\partial\over\partial\eta_n^A}.
\end{align}

Let us expand the superamplitide ${\mathscr A}_{n+1}$ in Grassmannian variables $\eta_s$
\begin{align}\label{}
  {\mathscr A}_{n+1} (\Phi_1,\ldots,\Phi_n, \Phi_s ) \,=\, &
  {\mathscr A}_{n+1} (\Phi_1,\ldots,\Phi_n, g_s^+ ) +
  \eta_s^A {\mathscr A}_{n+1} (\Phi_1,\ldots,\Phi_n, \Gamma_{sA} )
  \nonumber\\
  &+
  {1\over 2!} \eta_s^A \eta_s^B {\mathscr A}_{n+1} (\Phi_1,\ldots,\Phi_n, S_{sAB} ) +
  \cdots.
\end{align}
According to the degrees of the Grassmann odd $\eta_s$, we can express super soft theorem \eqref{soft-theorem-N=4-1} as following:
\begin{align}
  {\mathscr A}_{n+1}\big(\ldots, g_s^+\big) (\epsilon) \,&=\,
  \left( {1\over\epsilon^2} S_{\rm YM}^{(0)} + {1\over\epsilon} S_{\rm YM}^{(1)}\right)
  {\mathscr A}_n \,+\, {\cal O}(\epsilon^0),
  \\
  \label{soft-gluino-th-N=4-1}
  {\mathscr A}_{n+1}\big(\ldots, \Gamma_{sA} \big) (\epsilon)\,&=\,
  {1\over\epsilon} {\cal F}^{(0)}_A {\mathscr A}_n \,+\, {\cal O}(\epsilon^0),
  \\
  \label{soft-scalar-th-N=4-1}
  {\mathscr A}_{n+1}\big(\ldots, S_{sAB} \big) (\epsilon) \,&=\,
  {0\over\epsilon} \,+\, {\cal O}(\epsilon^0).
\end{align}
In the last equation, the term $0\over\epsilon$ implies that there is no singular term.
The soft gluon operators in ${\cal N}=4$ SYM are identical to the ones in pure Yang-Mills.
As mentioned in section \ref{Sec-Review}, the sub-leading soft gluon divergence is also vanishing in ${\cal N}=4$ SYM.
As we expected, the amplitudes involve more types of particle, including gluon, gluino and scalar in ${\cal N}=4$ SYM.
More interestingly, we find the soft divergence of amplitudes involving a soft fermionic gluino.
Notice that the leading soft gluino operator $\displaystyle {\cal F}^{(0)}_A$ involves the first order derivative with respect to the Grassmannian variables $\eta_1$ and $\eta_n$.
In fact, these two terms of $\displaystyle {\cal F}^{(0)}_A$ change helicity of corresponding external leg respectively.
And this preserves the total helicity as well as ${\rm SU}(4)$ R-symmetry before and after soft gluino emission.
We also provide an alternative derivation of the soft theorem in ${\cal N}=4$ SYM in appendix \ref{Sec-N=4-alternative}.
In the next subsection, we will check soft theorem by some examples in detail.
We will pay special attention to soft gluino theorem.


\subsection{MHV and NMHV Examples}
In the remainder of this section, we verify the soft theorem presented above by some examples in detail.
We take special care of the property of amplitudes when a gluino leg becomes soft.

The simplest example is MHV sector.
In this sector, one can study the amplitudes involving an arbitrary number of external legs.
In the holomorphic soft limit $\lambda_s\to\epsilon\lambda_s$, the soft theorem of pure gluonic MHV amplitudes gives
\begin{align}\label{MHV-gluonic-am-soft-1}
  {\cal A}_{n+1}^{\rm MHV}(\epsilon) \,&=\,{1\over\epsilon^2} S_{\rm YM}^{(0)} {\cal A}_n^{\rm MHV}
\end{align}
which is exact in $\epsilon$.

Now we study the amplitude $\displaystyle {\cal A}(\bar\Gamma^A, g^+,\ldots,g^+, g^-,\Gamma_B)$ involving a gluino-anti-gluino pair.
Using the soft gluino theorem \eqref{soft-gluino-th-N=4-1}, we get
\begin{align}\label{MHV-gluino-Example-1}
  {\cal A}_{n+1} \big(\bar\Gamma^A, g^+, \ldots, g^+, g^-, \Gamma_B \big) \,&=\,
  \delta^A_B
  {1\over\epsilon} {1\over\braket{s,1}} {\cal A}_n\big(g^-, g^+,\ldots,g^+, g^-\big)
  + {\cal O}(\epsilon^0).
\end{align}
in the holomorphic soft limit $\lambda_s\to\epsilon\lambda_s$.
We will show that there is no ${\cal O}(\epsilon^0)$ corrections in above relation.
Noting the following supersymmetric Ward identity (SWI) \cite{Dixon-Review-96,Elvang-Huang-1308.1697}:
\begin{align}
  {\cal A}_{n+1} \big(\bar\Gamma^A, g^+,\ldots, g^+, g_n^-, \Gamma_B \big) \,&=\,
  \delta^A_B
  {\braket{n,s}\over\braket{n,1}} {\cal A}_{n+1}\big(g^-, g^+,\ldots, g^+, g^-, g^+\big)
\end{align}
then using the soft gluon theorem \eqref{MHV-gluonic-am-soft-1}, we have
\begin{align}
  \delta^A_B
  {\epsilon \braket{n,s}\over\braket{n,1}} {\cal A}_{n+1}\big(g^-, g^+,\ldots, g^+, g^-, g^+\big) \,&=\,
  \delta^A_B
  {\epsilon \braket{n,s}\over\braket{n,1}} \times
  {1\over\epsilon^2} {\braket{n,1}\over\braket{n,s}\braket{s,1}}
  {\cal A}_n\big(g^-,g^+,\ldots,g^+, g^-\big)
  \nonumber\\
  &=\, \delta^A_B {1\over\epsilon} {1\over\braket{s,1}} {\cal A}_n\big(g^-,g^+,\ldots,g^+, g^-\big)
\end{align}
in the holomorphic soft limit $\lambda_s\to\epsilon\lambda_s$. This agrees with eq.~\eqref{MHV-gluino-Example-1}.

In the MHV sector, another a SWI involving two scalars is \cite{Dixon-Review-96,Elvang-Huang-1308.1697}:
\begin{align}
  {\cal A}_{n+1} \big(S_{12},g^-,g^+,\ldots,g^+, S_{34} \big) \,&=\,
  {\braket{2,s}^2\over\braket{2,1}^2} {\cal A}_{n+1}\big(g^-,g^-,g^+,\ldots,g^+, g^+\big).
\end{align}
Using the soft gluon theorem \eqref{MHV-gluonic-am-soft-1} for the right hand side of above equation, one finds that
\begin{align}
  {\cal A}_{n+1} \big(S_{12},g^-,g^+,\ldots,g^+, S_{34} \big)
  \,\xlongequal{\lambda_s\to\epsilon\lambda_s}&\,
  {\braket{2,s}^2\over\braket{2,1}^2} S^{(0)}_{\rm YM}{\cal A}_n\big(g^-,g^-,g^+,\ldots,g^+\big)
  \nonumber\\
  =&\,
  {\cal A}_{n+1} \big(S_{12},g^-,g^+,\ldots,g^+, S_{34} \big)
  \,\sim\, {\cal O}(\epsilon^0).
\end{align}
It agrees with the soft theorem \eqref{soft-scalar-th-N=4-1}. This also shows that the MHV amplitudes involving two scalars remain invariant under rescaling of momentum of one of scalars.

\subsection*{\it Six-point NMHV 2-gluino amplitudes}
Next we turn to the Next-to-MHV (NMHV) sector.
In this sector, it is difficult to check the amplitudes which consist of an arbitrary number of external legs.
Here we mainly check 6-point NMHV amplitudes involving a gluino-anti-gluino pair which were obtained by Feynman diagrams \cite{Kunszt-NPB1985}, also by solving supersymmetry Ward identity \cite{Bidder-Dunbar-Perkins-0505249} and BCFW recursion \cite{Luo-Wen-0501121}.

The first example is $\displaystyle A_6 (g_1^-, g_2^-, \bar\Gamma_3^A, \Gamma_{4B}, g_5^+,g_6^+ )$:
\begin{align}\label{}
  A_6\big(g_1^-, g_2^-, \bar\Gamma_3^A, \Gamma_{4B}, g_5^+,g_6^+\big) \,=\,&
  - {[4| 2+3 \ket{1}^2 [3|2+4\ket{1} \over s_{234}[23][34] \braket{56} \braket{61} [2| 3+4\ket{5}}\delta^A_B
  \nonumber\\
  &+ {[6| 1+2 \ket{3}^2 [6|1+2\ket{4} \over s_{612}[61][12] \braket{34} \braket{45} [2| 6+1\ket{5}}\delta^A_B.
\end{align}
In the soft limit $\lambda_4\to\epsilon\lambda_4$,
\begin{align}\label{}
  A_6\big(g_1^-, g_2^-, &\bar\Gamma_3^A, \Gamma_{4B}, g_5^+,g_6^+\big) \,=\,
  \delta^A_B {1\over\epsilon}
  {[65]^2\braket{53}^2 [6| 3+5 \ket{4} \over s_{35} [61][12] \braket{34} \braket{45} [23]\braket{35}}
  + {\cal O}(\epsilon^0)
  \nonumber\\
  &=\, \delta^A_B {1\over\epsilon}\left(
  {1 \over \braket{34}} A_5\big(g_1^-, g_2^-, g_3^-, g_5^+,g_6^+\big) \,+\,
  {1 \over \braket{45}} A_5\big(g_1^-, g_2^-, \bar\Gamma_3^C, \Gamma_{5C}, g_6^+\big)
  \right) + {\cal O}(\epsilon^0).
\end{align}
This agrees completely with the soft gluino theorem \eqref{soft-gluino-th-N=4-1}.
Here we have used two 5-point amplitudes follows:
\begin{align}\label{}
  A_5\big(g_1^-, g_2^-, g_3^-, g_5^+,g_6^+\big) \,&=\,
  {[56]^4 \over [12][23][35][56][61]},
 \\
  A_5\big(g_1^-, g_2^-, \bar\Gamma_3^A, \Gamma_{5B}, g_6^+\big) \,&=\,
  {[65]^2 [63] \over [12][23][35][61]}\delta^A_B.
\end{align}

The second example is the amplitude:
\begin{align}\label{}
  A_6\big(g_1^-, \bar\Gamma_2^A, g_3^-, \Gamma_{4B}, g_5^+,g_6^+\big) \,=\,&
  - {[4|2+3\ket{1}^2 [2| 3+4 \ket{1} \over s_{234} [23][34] \braket{56} \braket{61} [2|3+4\ket{5}} \delta^A_B
  \nonumber\\
  &+ {[6|1+2\ket{3}^2 [26] \braket{34} \over s_{612} [61][12] \braket{34} \braket{45} [2|6+1\ket{5}} \delta^A_B.
\end{align}
In the soft limit $\lambda_4\to\epsilon\lambda_4$, we have
\begin{align}\label{}
  A_6\big(g_1^-, \bar\Gamma_2^A, g_3^-, \Gamma_{4B}, g_5^+,g_6^+\big) \,&=\,
  \delta^A_B {1\over\epsilon}
  {[65]^2\braket{53}^2 [26] \over s_{35} [61][12] \braket{45} [23]\braket{35}}
  + {\cal O}(\epsilon^0)
  \nonumber\\
  &=\,\delta^A_B {1\over\epsilon} {1\over\braket{45}}
  A_5\big(g_1^-, \bar\Gamma_2^A, g_3^-, \Gamma_{5A}, g_6^+\big)
  + {\cal O}(\epsilon^0)
\end{align}
where
\begin{align}\label{}
  A_5\big(g_1^-, \bar\Gamma_2^A, g_3^-, \Gamma_{5B}, g_6^+\big) \,&=\,
  {[56] \over [26]}  {[56]^4 \over [12] [23] [35] [56] [61]} \delta^A_B.
\end{align}
This also agrees with the soft gluino theorem \eqref{soft-gluino-th-N=4-1}.

Similarly, after some calculation we have
\begin{align}\label{}
  A_6\big(\bar\Gamma_1^{A}, g_2^-, g_3^-, \Gamma_{4B}, g_5^+,g_6^+\big) \,\xlongequal{\lambda_4\to\epsilon\lambda_4}\,&
  {1\over\epsilon} {1\over\braket{45}}
  A_5\big(\bar\Gamma_1^{A}, g_2^-, g_3^-, \Gamma_{5B}, g_6^+\big)
  + {\cal O}(\epsilon^0),
  \\
  A_6\big(g_1^-, \bar\Gamma_2^{A}, g_3^-, g_4^+, \Gamma_{5B}, g_6^+\big) \,\xlongequal{\lambda_5\to\epsilon\lambda_5}\,&
  {1\over\epsilon} {1\over\braket{45}}
  A_5\big(g_1^-, \bar\Gamma_2^{A}, g_3^-, \Gamma_{4B}, g_6^+\big)
  \nonumber\\
  &+ {1\over\epsilon}{1\over\braket{56}}
  A_5\big(g_1^-, \bar\Gamma_2^{A}, g_3^-, g_4^+, \Gamma_{6B}\big)
  + {\cal O}(\epsilon^0),
  \\
  A_6\big(\bar\Gamma_1^{A}, g_2^-, g_3^-, g_4^+, \Gamma_{5B}, g_6^+\big) \,\xlongequal{\lambda_5\to\epsilon\lambda_5}\,&
  {1\over\epsilon} {1\over\braket{45}}
  A_5\big(\bar\Gamma_1^{A}, g_2^-, g_3^-, \Gamma_{4B}, g_6^+\big)
  \nonumber\\
  &+ {1\over\epsilon} {1\over\braket{56}}
  A_5\big(\bar\Gamma_1^{A}, g_2^-, g_3^-, g_4^+, \Gamma_{6B}\big)
  + {\cal O}(\epsilon^0),
  \\
  A_6\big(\bar\Gamma_1^{A}, g_2^-, \Gamma_{3B}, g_4^-, g_5^+, g_6^+\big) \,\xlongequal{\lambda_3\to\epsilon\lambda_3}\,&
  {0\over\epsilon} + {\cal O}(\epsilon^0),
  \\
  A_6\big(g_1^-, \bar\Gamma_2^{A}, \Gamma_{3B}, g_4^-, g_5^+, g_6^+\big) \,\xlongequal{\lambda_3\to\epsilon\lambda_3}\,&
  {1\over\epsilon} {1\over\braket{23}} \delta_B^A\,A_5\big(g_1^-, g_2^-, g_4^-, g_5^+, g_6^+\big)
  + {\cal O}(\epsilon^0),
  \\
  A_6\big(g_1^-, g_2^-, \Gamma_{3B}, \bar\Gamma_4^{A}, g_5^+, g_6^+\big) \,\xlongequal{\lambda_3\to\epsilon\lambda_3}\,&
  {1\over\epsilon} {1\over\braket{34}} \delta_B^A\,
  A_5\big(g_1^-, g_2^-, g_4^-, g_5^+, g_6^+\big)
  + {\cal O}(\epsilon^0),
  \\
  %
  %
  A_6\big(\bar\Gamma_1^{A}, \Gamma_{2B}, g_3^-, g_4^+, g_5^-, g_6^+\big) \,\xlongequal{\lambda_2\to\epsilon\lambda_2}\,&
  {1\over\epsilon} {1\over\braket{12}} \delta_B^A\,
  A_5\big(g_1^-, g_3^-, g_4^+, g_5^-, g_6^+\big)
  + {\cal O}(\epsilon^0),
  \\
  A_6\big(g_1^-, \Gamma_{2B}, \bar\Gamma_3^{A}, g_4^+, g_5^-, g_6^+\big) \,\xlongequal{\lambda_2\to\epsilon\lambda_2}\,&
  {1\over\epsilon} {1\over\braket{23}} \delta_B^A\,
  A_5\big(g_1^-, g_3^-, g_4^+, g_5^-, g_6^+\big)
  + {\cal O}(\epsilon^0),
  \\
  A_6\big(g_1^-, \Gamma_{2B}, g_3^-, g_4^+, \bar\Gamma_5^{A}, g_6^+\big) \,\xlongequal{\lambda_2\to\epsilon\lambda_2}\,&
  {0\over\epsilon} + {\cal O}(\epsilon^0).
\end{align}
Here we omit some details.
These examples strongly support the soft gluino theorem \eqref{soft-gluino-th-N=4-1}.
In addition, the same checks can be done for other six-point NMHV amplitudes in ${\cal N}=4$ SYM, for example tree-level four-gluino two-gluon amplitudes and six-gluino amplitudes \cite{Luo-Wen-0502009}.

\section{Super soft theorem in ${\cal N}=8$ supergravity} \label{Sec-N=8}
The ${\cal N}=8$ Supergravity is the most symmetric quantum field theory in 4 dimensions.
In this section, we study super soft theorem in ${\cal N}=8$ SUGRA.
First we derive the soft theorem by using super-BCFW recursions at tree level.
Then we verify soft divergences of scattering amplitudes of ${\cal N}=8$ SUGRA, in particular soft gravitino and soft gravi-photon divergences, by some MHV tree-level amplitudes exactly.
We also give ``KLT-like relations'' between soft operators in ${\cal N}=8$ SUGRA and ones in ${\cal N}=4$ SYM at the end of this section.


The ${\cal N}=8$ SUGRA consists of $256$ massless on-shell fields which can be characterized as
\begin{align}\label{N=8-supermultiplet-as-square-of-N=4}
  ({\cal N}=8~\text{SUGRA}) ~\sim~ ({\cal N}=4~\text{SYM}) \otimes ({\cal N}=4~\text{SYM}).
\end{align}
These on-shell  fields form a CPT-self-conjugate supermultiplet and may be organized into a single on-shell superfield $\Phi$.
With the help of the Grassmann odd variables $\eta^A$, one can expand on-shell superfield $\Phi$ as following:
\begin{align}\label{}
  \Phi(p,\eta) \,=\,& h^+(p) + \eta^A \psi_A(p)
  + {1\over 2!} \eta^A \eta^B v_{AB}(p)
  + {1\over 3!} \eta^A \eta^B \eta^C \chi_{ABC}(p)
  \nonumber\\
  &+ {1\over 4!} \eta^A\eta^B\eta^C\eta^D S_{ABCD}(p) + \cdots +
  \eta^{1}\eta^{2}\eta^{3}\eta^{4}\eta^{5}\eta^{6}\eta^{7}\eta^{8} h^-(p).
\end{align}
Here $A, B, \ldots = 1,2,\ldots, {\cal N}$ are ${\rm SU}(8)$ R-symmetry indices and each state above is fully antisymmetric in these labels.

In ${\cal N}=8$ on-shell superspace, there are also fundamental three-point superamplitudes:
\begin{align}\label{}
  {\mathscr M}_3^{\rm MHV} \,&=\, {\displaystyle \delta^4(p) \over
  \big( \braket{12}\braket{23}\braket{31} \big)^2 }
  \delta^{16}\big(\lambda_1^\alpha\eta_1^A + \lambda_2^\alpha\eta_2^A + \lambda_3^\alpha\eta_3^A \big), \\
  {\mathscr M}_3^{\overline{\rm MHV}} \,&=\,
  {\displaystyle \delta^4(p) \over  \big( [12][23][31] \big)^2 }
  \delta^8\big([12] \eta_3^A + [23] \eta_1^A + [31] \eta_2^A \big).
\end{align}
Here each is just the square of corresponding three-point superamplitude of ${\cal N}=4$ SYM.

\subsection{Super soft theorem in ${\cal N}=8$ SUGRA}
Now we start to derive the soft theorem.
Consider an on-shell $(n+1)$-point superamplitudes in ${\cal N}=8$ SUGRA with a soft external leg $\Phi_s(\lambda_s,\tilde\lambda_s,\eta_s)$\footnote{
Here soft particle may be any one in ${\cal N}=8$ supermultiplet \eqref{N=8-supermultiplet-as-square-of-N=4}, including graviton (spin-2), gravitino (spin-3/2), gravi-photon (spin-1), gravi-photino (spin-1/2) and scalar (spin-0).
}
\begin{align}\label{}
  {\mathscr M}_{n+1} \,\equiv\,
  {\mathscr M}_{n+1}\big(\{\lambda_1,\tilde\lambda_1, \eta_1\}, \cdots,\{\lambda_n,\tilde\lambda_n, \eta_n\}, \{\lambda_s,\tilde\lambda_s, \eta_s\} \big).
\end{align}
Let us choose the following super-shift:
\begin{align}\label{}
  \lambda_s(z)=\lambda_s+z\lambda_n,\quad \tilde\lambda_n(z)=\tilde\lambda_n-z\tilde\lambda_s,\quad \eta_n(z)=\eta_n - z\eta_s.
\end{align}
Using the analysis similar to SYM, the super-BCFW recursion gives
\begin{align}\label{}
  {\mathscr M}_{n+1} \,=\,& \sum_{a=1}^{n-1} {[s,a] \braket{n,a}^2 \over \braket{s,a} \braket{n,s}^2 }
  \nonumber\\
  &\times
  {\mathscr M}_n \Big( \ldots,
  \{\lambda_a, \tilde\lambda_a + {\braket{n,s} \over \braket{n,a}}\tilde\lambda_s,
  \eta_a + {\braket{n,s} \over \braket{n,a}} \eta_s \}, \ldots,
  \{\lambda_n, \tilde\lambda_n + {\braket{s,a} \over \braket{n,a}}\tilde\lambda_s,
  \eta_n + {\braket{s,a} \over \braket{n,a}}\eta_s \} \Big).
\end{align}
Here one has omitted terms which stay finite in the holomorphic soft limits \eqref{holomorphic-soft-limit-1}.
Applying the deformation $\lambda_s \to \epsilon\lambda_s$ to above formula, one gets
\begin{align}\label{soft-theorem-N=8-soft-limit-1}
  {\mathscr M}_{n+1} (\epsilon) &\,=\,
  {1\over \epsilon^3}\sum_{a=1}^{n-1} {[s,a] \braket{n,a}^2 \over \braket{s,a} \braket{n,s}^2 }
  \nonumber\\
  &\times
  {\mathscr M}_n \Big( \ldots,
  \{\lambda_a, \tilde\lambda_a + \epsilon {\braket{n,s} \over \braket{n,a}}\tilde\lambda_s,
  \eta_a + \epsilon {\braket{n,s} \over \braket{n,a}} \eta_s \}, \ldots,
  \{\lambda_n, \tilde\lambda_n + \epsilon {\braket{s,a} \over \braket{n,a}}\tilde\lambda_s,
  \eta_n + \epsilon {\braket{s,a} \over \braket{n,a}}\eta_s \} \Big).
\end{align}
Performing Taylor expansion of ${\cal M}(\epsilon)$ around $\epsilon=0$, one obtains the super soft theorem:
\begin{align}\label{soft-theorem-N=8-1}
  {\mathscr M}_{n+1} (\epsilon) &=
  \left(
  {1\over \epsilon^3}{\cal S}^{(0)} + {1\over \epsilon^2}{\cal S}^{(1)} + {1\over \epsilon}{\cal S}^{(2)}
  \right)
  {\mathscr M}_n \,+\, {\cal O}(\epsilon^0).
\end{align}
Here the leading soft factor is same with the one in non-supersymmetric gravity theory,
\begin{align}\label{}
  {\cal S}^{(0)} \,&=\, \sum_{a=1}^{n-1} {[s,a] \braket{n, a}^2 \over \braket{s,a}\braket{n, s}^2}
  \,=\, S^{(0)}.
\end{align}
The sub-leading soft operator consists of two parts:
\begin{align}\label{}
  {\cal S}^{(1)} \,&=\, \sum_{a=1}^{n-1} {[s,a] \braket{n,a} \over \braket{s,a} \braket{n,s} }
  \left(
  \tilde\lambda_{s\dot\alpha} {\partial\over\partial\tilde\lambda_{a\dot\alpha} } +
  \eta_s^A {\partial\over\partial\eta_a^A}
  \right)
  \,=\, S^{(1)} \,+\, \eta_s^A {\cal S}^{(1)}_A,
\end{align}
while the sub-sub-leading soft operator consists of three parts:
\begin{align}\label{}
  {\cal S}^{(2)} \,&=\, S^{(2)} \,+\, \eta_s^A {\cal S}^{(2)}_A
  \,+\, \frac12\eta_s^A\eta_s^B {\cal S}^{(2)}_{AB},
  \\
  S^{(2)} \,&=\, \frac12 \sum_{a=1}^{n} {[s,a] \over \braket{s,a}}
  \tilde\lambda_{s\dot\alpha}\tilde\lambda_{s\dot\beta}
  {\partial^2\over\partial\tilde\lambda_{a\dot\alpha}\partial\tilde\lambda_{a\dot\beta}},
  \\
  {\cal S}^{(2)}_A \,&=\, \sum_{a=1}^{n} {[s,a] \over \braket{s,a}}
  \tilde\lambda_{s\dot\alpha} {\partial^2\over\partial\tilde\lambda_{a\dot\alpha}\partial\eta_a^A},
  \\
  {\cal S}^{(2)}_{AB} \,&=\, \sum_{a=1}^{n} {[s,a] \over \braket{s,a}}
  {\partial^2\over\partial\eta_a^B\partial\eta_a^A}.
\end{align}
See Appendix \ref{Sec-N=8-NNLO-Soft-Operators} for some calculational details.
Expanding superamplitude ${\mathscr M}_{n+1}$ in Grassman odd variables $\eta_s$, we have
\begin{align}\label{}
  {\mathscr M}_{n+1} \big(\Phi_1, \ldots, \Phi_n, \Phi_s\big) \,=\, &
  {\mathscr M}_{n+1} \big(\Phi_1, \ldots, \Phi_n, h_s^+\big)
  + \eta_s^A {\mathscr M}_{n+1} \big(\Phi_1, \ldots, \Phi_n, \psi_{sA}\big)
  \nonumber\\
  &+ {1\over2}\eta_s^A\eta_s^B {\mathscr M}_{n+1} \big(\Phi_1, \ldots, \Phi_n, v_{sAB}\big)
  + \cdots.
\end{align}
Thus we can express the soft theorem \eqref{soft-theorem-N=8-1} in ${\cal N}=8$ SUGRA as
\begin{align}
  \label{N=8-soft-graviton-1}
  \text{\it soft graviton:}\quad
  {\mathscr M}_{n+1} \big(\ldots, h_s^+\big) (\epsilon) \,&=\,
  \left(
  {1\over \epsilon^3}{S}^{(0)} + {1\over \epsilon^2}{S}^{(1)} + {1\over \epsilon}{S}^{(2)}
  \right)
  {\mathscr M}_n + {\cal O}(\epsilon^0),
  \\
  \label{N=8-soft-gravitino-1}
  \text{\it soft gravitino:}\quad
  {\mathscr M}_{n+1} \big(\ldots, \psi_{sA}\big) (\epsilon) \,&=\,
  \left(
  {1\over \epsilon^2}{\cal S}^{(1)}_A + {1\over \epsilon}{\cal S}^{(2)}_A
  \right)
  {\mathscr M}_n + {\cal O}(\epsilon^0),
  \\
  \label{N=8-soft-gravi_photon-1}
  \text{\it soft gravi-photon:}\quad
  {\mathscr M}_{n+1} \big(\ldots, v_{sAB}\big) (\epsilon) \,&=\,
  {1\over \epsilon} {\cal S}^{(2)}_{AB} {\mathscr M}_n + {\cal O}(\epsilon^0),
  \\
  \label{N=8-soft-gravi_photino-1}
  \text{\it soft gravi-photino:}\quad
  {\mathscr M}_{n+1} \big(\ldots, \chi_{sABC}\big) (\epsilon) \,&=\, {0\over \epsilon} + {\cal O}(\epsilon^0),
  \\
  \label{N=8-soft-scalar-1}
  \text{\it soft scalar:}\quad
  {\mathscr M}_{n+1} \big(\ldots, S_{sABCD}\big) (\epsilon) \,&=\, {0\over \epsilon} + {\cal O}(\epsilon^0).
\end{align}
There are more contents in ${\cal N}=8$ SUGRA.
Scattering amplitudes of ${\cal N}=8$ SUGRA involve more types of particle.
That is, every (hard or soft) external leg in amplitudes may be any particle of 4D ${\cal N}=8$ SUGRA.
Thus the soft graviton theorem \eqref{N=8-soft-graviton-1} in SUGRA incorporates the soft graviton theorem for pure graviton amplitudes.
Besides the soft graviton theorem, one obtains leading and sub-leading soft gravitino divergences and leading soft gravi-photon divergence. Also one finds that there are no soft gravi-photino divergence and soft scalar divergence.

In the next subsection, we will check soft theorems by some examples in the MHV sector of ${\cal N}=8$ SUGRA in detail.
We will pay special attention to the soft gravitino divergence and the soft gravi-photon divergence.

\subsection{MHV Examples}\label{Sec-N=8-Examples}
For the soft graviton theorem, in particular leading and sub-leading orders, there are a great deal of study and investigation on both theoretical derivations and special examples check so far \cite{Weinberg1965,Cachazo-Strominger-1404.4091,White-1103.2981}.
In this subsection, we check soft theorem of ${\cal N}=8$ SUGRA amplitudes by some examples of the MHV sector.
We mainly focus on the leading and the sub-leading soft gravitino divergences and the leading soft gravi-photon divergence, as well as the property of amplitudes with soft scalar.

First of all, we analyse a special class of amplitudes which are proportional to MHV amplitudes of gravitons.
For such amplitudes, we can check soft gravitino divergences or soft gravi-photon divergences by using only soft graviton theorem, eq.~\eqref{new-soft-gravitom-th-1} or eq.~\eqref{N=8-soft-graviton-1}.
In  MHV sector of ${\cal N}=8$ SUGRA, there exists the following supersymmetry Ward identities\footnote{
Here the generalized Kronecker delta-symbol is defined as
\begin{align*}
  \delta^{A_1 \dots A_n}_{B_1 \dots B_n} \,\equiv\,
  \sum_{\sigma \in S_n} \operatorname{sgn}(\sigma)\, \delta^{A_{\sigma(1)}}_{B_1}\cdots\delta^{A_{\sigma(n)}}_{B_n}.
\end{align*}
}:
\begin{align}
  \label{N=8-SWI-gravitino-1}
  {\cal M}_{n+1}\big(h_1^-,\psi_2^{A}, h_3^+, \ldots, h_n^+, \psi_{sB}\big) \,&=\,
  \delta_B^A\, {\braket{1,s}\over\braket{1,2}}\,
  {\cal M}_{n+1}\big(h_1^-, h_2^-, h_3^+, \ldots, h_n^+, h_s^+\big),
  \\
  %
  \label{N=8-SWI-gravi_photon-1}
  {\cal M}_{n+1}\big(h_1^-, v_2^{AB}, h_3^+, \ldots, h_n^+, v_{sCD}\big) \,&=\,
  \delta^{AB}_{CD}\, {\braket{1,s}^2\over\braket{1,2}^2}\,
  {\cal M}_{n+1}\big(h_1^-, h_2^-, h_3^+, \ldots, h_n^+, h_s^+\big),
  \\
  %
  \label{N=8-SWI-gravi_photino-1}
  {\cal M}_{n+1}\big(h_1^-, \chi_2^{ABC}, h_3^+, \ldots, h_n^+, \chi_{sDEF}\big) \,&=\,
  \delta^{ABC}_{DEF}\, {\braket{1,s}^3\over\braket{1,2}^3}\,
  {\cal M}_{n+1}\big(h_1^-, h_2^-, h_3^+, \ldots, h_n^+, h_s^+\big),
  \\
  %
  \label{N=8-SWI-scalar-1}
  {\cal M}_{n+1}\big(h_1^-, S_2^{ABCD}, h_3^+, \ldots, h_n^+, S_{sEFGH}\big) \,&=\,
  \delta^{ABCD}_{EFGH}\, {\braket{1,s}^4\over\braket{1,2}^4}
  {\cal M}_{n+1}\big(h_1^-, h_2^-, h_3^+, \ldots, h_n^+, h_s^+\big)
\end{align}
where $\displaystyle S^{ABCD}={1\over 4!}\epsilon^{ABCDEFGH} S_{EFGH}$.
With the help of these Ward identities, we can study soft divergences of amplitudes involving lower-spin soft particles by using only soft graviton theorem.

\subsection*{\it Soft gravitino}
Here we study the soft gravitino divergence.
For the right hand side of SWI \eqref{N=8-SWI-gravitino-1}, by using the soft graviton theorem we have
\begin{align}\label{N=8-SWI-gravitino-X1}
  \delta^A_B\,  {\braket{1,s}\over\braket{1,2}}
  \left({1\over\epsilon^2}S^{(0)} + {1\over\epsilon}S^{(1)}\right)
  {\cal M}_{n}\big(h_1^-,h_2^-,h_3^+,\ldots,h_n^+\big)
  + {\cal O}(\epsilon^0).
\end{align}
First we consider the leading order ${\cal O}(\epsilon^{-2})$:
\begin{align}\label{N=8-MHV-Ex-gravitino-L-1}
  \delta^A_B\, {1\over\epsilon^2} {\braket{1,s}\over\braket{1,2}} S^{(0)}
  {\cal M}_{n}\big(h_1^-,h_2^-,h_3^+,\ldots,h_n^+\big)
  \,=\,
  \delta^A_B\, {1\over\epsilon^2} \sum_{a=2}^{n-1}  {\braket{1,a}\over\braket{1,2}}
  {[s,a] \braket{n,a} \over \braket{s,a} \braket{n,s}}
  {\cal M}_{n}\big(h_1^-,h_2^-,h_3^+,\ldots,h_n^+\big).
\end{align}
In order to get the sub-leading order, substituting soft operator $S^{(1)}$ into the right hand side of eq.~\eqref{N=8-SWI-gravitino-X1}, we have
\begin{align}
  \delta^A_B\, {1\over\epsilon} {\braket{1,s}\over\braket{1,2}} S^{(1)}
  {\cal M}_{n}\big(h_1^-,h_2^-,h_3^+,\ldots,h_n^+\big)
  \,&=\,
  \delta^A_B\, {1\over\epsilon} \sum_{a=2}^{n} {[s,a] \over \braket{s,a}} {\braket{1,a} \over \braket{1,2}}
  \tilde\lambda_{s\dot\alpha} {\partial\over\partial\tilde\lambda_{a\dot\alpha}}
  {\cal M}_{n}\big(h_1^-,h_2^-,h_3^+,\ldots,h_n^+\big).
  \label{N=8-MHV-Ex-gravitino-NL-1}
\end{align}
Here the gauge freedom of $S^{(0)}$ is fixed by taking $x=y=1$.

Next we study the soft-gravitino divergence by using directly the soft gravitino theorem \eqref{N=8-soft-gravitino-1}.
From the left hand side of the identity \eqref{N=8-SWI-gravitino-1}, using the leading soft gravitino theorem \eqref{N=8-soft-gravitino-1}, one can obtain:
\begin{align}
  {1\over \epsilon^2} \sum_{a=1}^{n-1} & {[s,a] \braket{n,a} \over \braket{s,a} \braket{n,s} }
  {\cal M}_{n}\big(h_1^-,\psi_2^{A},\ldots,{\mathscr Q}_{aB}\Phi_a,\ldots,h_n^+\big)
  \nonumber\\
  =~&
  {1\over \epsilon^2} \Bigg(
  {[s,2] \braket{n,2} \over \braket{s,2} \braket{n,s} }
  {\cal M}_{n}\big(h_1^-, h_2^-, \ldots, h_n^+\big)\delta^A_B
  +
  \sum_{a=3}^{n-1} {[s,a] \braket{n,a} \over \braket{s,a} \braket{n,s} }
  {\cal M}_{n}\big(h_1^-,\psi_2^{A},\ldots,\psi_{aB},\ldots,h_n^+\big)
  \bigg)
  \nonumber\\
  =~&
  \delta^A_B\,
  {1\over \epsilon^2}  \sum_{a=2}^{n-1} {[s,a] \braket{n,a} \over \braket{s,a} \braket{n,s} }
  {\braket{1,a} \over \braket{1,2}}
  {\cal M}_{n}\big(h_1^-,h_2^-,h_3^+,\ldots,h_n^+\big).
  \label{N=8-MHV-Ex-gravitino-L-2}
\end{align}
This gives the same result as eq.~\eqref{N=8-MHV-Ex-gravitino-L-1}.
Here one has used the SUSY Ward identity \eqref{N=8-SWI-gravitino-1} and the operator ${\mathscr Q}_{aA}$ is defined by
\begin{align}
  {\mathscr Q}_{aA} h_a^+ \,\equiv\, \psi_{aA}, \quad
  {\mathscr Q}_{aA} \psi_{aB} \,\equiv\, v_{aAB},  \quad
  {\mathscr Q}_{aA} \psi_a^{B} \,\equiv\, \delta_A^B\, h_a^-, \quad
  {\mathscr Q}_{aA} h_a^- \,\equiv\, 0,  ~~  \cdots.
\end{align}
Roughly speaking, it make spin (or helicity) of the particle reduce by one half when an operator ${\mathscr Q}_{aA}$ act on this particle.
Similarly, applying straightforwardly the sub-leading soft gravitino theorem \eqref{N=8-soft-gravitino-1} to the left hand side of the identity \eqref{N=8-SWI-gravitino-1} gives
\begin{align}
  {1\over \epsilon} \sum_{a=1}^{n} {[s,a] \over \braket{s,a}} &
  \tilde\lambda_{s\dot\alpha} {\partial\over\partial\tilde\lambda_{a\dot\alpha}}
  {\cal M}_{n}\big(h_1^-,\psi_2^{A},\ldots,{\mathscr Q}_{aB}\Phi_a,\ldots,h_n^+\big)
  \nonumber\\
  \,&=\,
  \delta^A_B\,
  {1\over \epsilon} \sum_{a=2}^{n} {[s,a] \over \braket{s,a}}  {\braket{1,a}\over\braket{1,2}}
  \tilde\lambda_{s\dot\alpha} {\partial\over\partial\tilde\lambda_{a\dot\alpha}}
  {\cal M}_{n}\big(h_1^-,h_2^-,h_3^+,\ldots,h_n^+\big).
\end{align}
This give the same result as the one from the soft graviton theorem, eq.~\eqref{N=8-MHV-Ex-gravitino-NL-1}.

\subsection*{\it Soft gravi-photon}
Since there is only leading soft gravi-photon divergence, just Weinberg's leading soft graviton theorem is need.
In the holomorphic soft limit $\lambda_s\to\epsilon\lambda_s$, using leading soft graviton theorem\footnote{
It is necessary to notice that gauge freedoms in soft factor $S^{(0)}$ are fixed by setting $x=y=1$.
} the right hand side of the SWI \eqref{N=8-SWI-gravi_photon-1} becomes
\begin{align}\label{}
  {1\over\epsilon}\, \delta^{AB}_{CD}\, {\braket{1,s}^2\over\braket{1,2}^2} &
  S^{(0)} {\cal M}_{n}\big(h_1^-, h_2^-, h_3^+, \ldots, h_n^+\big) \,+\, {\cal O}(\epsilon^0)
  \nonumber\\
  &=\,
  {1\over\epsilon}\, \delta^{AB}_{CD}\,
  \sum_{a=2}^{n} {\braket{1,a}^2\over\braket{1,2}^2} {[s,a] \over \braket{s,a}}
  {\cal M}_{n}\big(h_1^-, h_2^-, h_3^+, \ldots, h_n^+\big) \,+\, {\cal O}(\epsilon^0).
\end{align}
On the other hand, by applying straightforwardly the soft gravi-photon theorem \eqref{N=8-soft-gravi_photon-1} to the left hand side of the SWI \eqref{N=8-SWI-gravi_photon-1}, one gets
\begin{align}\label{}
  {\cal M}_{n+1} & \big(h_1^-, v_2^{AB}, \ldots, v_{s\,CD}\big) \,=\,
  {1\over\epsilon}  \sum_{a=1}^{n} {[s,a] \over \braket{s,a}}
  {\cal M}_{n}\big(h_1^-, v_2^{AB-}, \ldots, {\mathscr Q}_{aD}{\mathscr Q}_{aC}\Phi_a,\ldots, h_n^+\big)
  + {\cal O}(\epsilon^0)
  \nonumber\\
  &=\,
  {1\over\epsilon} \Bigg(
  {[s,2] \over \braket{s,2}}
  {\cal M}_{n}\big(h_1^-, h_2^-, \ldots, h_n^+\big) \delta^{AB}_{CD}
  +
  \sum_{a=3}^{n} {[s,a] \over \braket{s,a}}
  {\cal M}_{n}\big(h_1^-, v_2^{AB}, \ldots, v_{aCD}, \ldots, h_n^+\big)
  \Bigg)
  + {\cal O}(\epsilon^0)
  \nonumber\\
  &=\,
  {1\over\epsilon}\, \delta^{AB}_{CD}
  \sum_{a=2}^{n} {[s,a] \over \braket{s,a}}  {\braket{1,a}^2\over\braket{1,2}^2}
  {\cal M}_{n}\big(h_1^-, h_2^-, h_3^+, \ldots, h_n^+\big) \,+\, {\cal O}(\epsilon^0).
\end{align}
Here one has used the Ward identity \eqref{N=8-SWI-gravi_photon-1}.

Next we analyse two 4-point amplitudes which have been computed by using both generating function method proposed in \cite{Bianchi-Elvang-Freedman-0805.0757} and Feynman diagrams respectively in \cite{Kallosh-Lee-Rube-0811.3417}.
The first example is {\it 4-gravi-photon amplitude}:
\begin{align}\label{}
  {\cal M}_4 \big(v^{AB}, v^{CD}, v_{EF}, v_{GH}\big) \,&=\,
  \braket{12}^2 [34]^2  \left(
  {1\over t}\delta_{EF}^{AB}\delta_{GH}^{CD}
  + {1\over u}\delta_{GH}^{AB}\delta_{EF}^{CD}
  + {1\over s}\delta_{EFGH}^{ABCD}
  \right)
\end{align}
where $s$, $t$, $u$ are Mandelstam variables
\begin{align}\label{}
  s \,&=\, (k_1 + k_2)^2 \,=\, (k_3 + k_4)^2  \,=\, s_{12} \,=\, s_{34}, \\
  t \,&=\, (k_1 + k_3)^2 \,=\, (k_2 + k_4)^2  \,=\, s_{13} \,=\, s_{24}, \\
  u \,&=\, (k_1 + k_4)^2 \,=\, (k_2 + k_3)^2  \,=\, s_{14} \,=\, s_{23}.
\end{align}
In holomorphic soft limit $\lambda_4\to\epsilon\lambda_4$, the amplitude becomes
\begin{align}\label{}
  {\cal M}_4 \big(v^{AB}, v^{CD}, v_{EF}, \epsilon\, v_{GH}\big)
  \,&=\,
  {1\over\epsilon}  \left(
  {\braket{12}^2 [34]^2 \over \braket{24}[24]}\delta_{EF}^{AB}\delta_{GH}^{CD}
  + {\braket{12}^2 [34]^2 \over \braket{14}[14]}\delta_{GH}^{AB}\delta_{EF}^{CD}
  + {\braket{12}^2 [34]^2 \over \braket{34}[34]}\delta_{EFGH}^{ABCD}
  \right)
  \nonumber\\
  \,&=\,
  {1\over\epsilon}  \left(
  {[24] \over \braket{24}}
  {\braket{12}^4 \over \braket{13}^2} \delta_{EF}^{AB}\delta_{GH}^{CD}
  + {[14] \over \braket{14}}
  {\braket{12}^4 \over \braket{23}^2} \delta_{GH}^{AB}\delta_{EF}^{CD}
  + { [34] \braket{12}^2 \over \braket{34}} \delta_{EFGH}^{ABCD}
  \right)
\end{align}
which is exact in $\epsilon$.
In the last line, one used the momentum conservation conditions: $\displaystyle \braket{12}[24]=-\braket{13}[34]$ and $\displaystyle \braket{21}[14]=-\braket{23}[34]$.

Notice that following 3-point amplitudes of ${\cal N}=8$ SUGRA:
\begin{align}\label{}
  {\cal M}_3 \big(h_1^-, v_2^{CD}, v_{3EF}\big)
  \,&=\, {\braket{12}^4 \over \braket{23}^2 } \delta^{CD}_{EF},
  \\
  {\cal M}_3 \big(v_1^{AB}, h_2^-, v_{3EF} \big)
  \,&=\, {\braket{12}^4 \over \braket{31}^2 } \delta^{AB}_{EF},
  \\
  {\cal M}_3 \big(v_1^{AB}, v_2^{CD}, S_{3EFGH}\big)
  \,&=\, \braket{12}^2 \delta^{ABCD}_{EFGH}.
\end{align}
Then one finds that
\begin{align}\label{}
  {\cal M}_4 \big(v^{AB}, v^{CD}, v_{EF}, \epsilon v_{GH}\big)
  \,=\,
  {1\over\epsilon}  \Bigg(&
  {[24] \over \braket{24}}
  {\cal M}_3 \big(v_1^{AB}, h_2^-, v_{3EF} \big) \delta_{GH}^{CD}
  + {[14] \over \braket{14}}
  {\cal M}_3 \big(h_1^-, v_2^{CD}, v_{3EF}\big) \delta_{GH}^{AB}
  \nonumber\\
  &+ {[34] \over \braket{34}}
  {\cal M}_3 \big(v_1^{AB}, v_2^{CD}, S_{3EFGH}\big)
  \Bigg).
\end{align}
Very nice! This is just the result from soft gravi-photon theorem.

Another example is {\it 2-scalar 2-gravi-photon amplitude}:
\begin{align}\label{2-scalar-2-gravi-photon-amplitude}
  {\cal M}_4 \big(v^{AB}, & v_{CD}, S_{EFGH}, S_{IJKL}\big) \nonumber\\
  =\,& \braket{13}^2[23]^2
  \bigg(
  {1 \over s}\delta^{AB}_{CD}\epsilon_{EFGHIJKL}
  + {3! \over t}\delta^{AB}_{[EF} \epsilon_{GH] IJKLCD}
  + {3! \over u}\delta^{AB}_{[IJ} \epsilon_{KL] EFGHCD}
  \bigg).
\end{align}
We study the property of this amplitude when external particle $v_{CD}$ becomes soft.
In holomorphic soft limit $\lambda_2\to\epsilon\lambda_2$, this amplitude becomes
\begin{align}\label{}
  {\cal M}_4 \big(v^{AB}, \epsilon\,v_{CD}, S_{EFGH}, S_{IJKL}\big) \,=\,
  {1 \over \epsilon}
  \bigg( &
  {[12] \over \braket{12}}  {\braket{13}^2 \braket{14}^2 \over \braket{34}^2}
  \delta^{AB}_{CD}\epsilon_{EFGHIJKL}
  + 3!{[24] \over \braket{24}}  \braket{14}^2
  \delta^{AB}_{[EF} \epsilon_{GH] IJKLCD}
  \nonumber\\
  &
  + 3!{[23] \over \braket{23}} \braket{13}^2  \delta^{AB}_{[IJ} \epsilon_{KL] EFGHCD}
  \bigg).
\end{align}
Here used the momentum conservation conditions: $\displaystyle {[23] \over [12]} = {\braket{14} \over \braket{34}}$ and $\displaystyle \braket{13}[23]=-\braket{14}[24]$.

Noting the following three-point MHV amplitudes:
\begin{align}\label{}
  {\cal M}_3 \big(h_1^-, S_{3EFGH}, S_{4IJKL}\big) \,&=\,
  {\braket{13}^2 \braket{14}^2 \over \braket{34}^2}
  \epsilon_{EFGHIJKL}
  \\
  {\cal M}_3 \big(v_1^{AB}, v_3^{MN}, S_{4IJKL}\big) \,&=\, \braket{13}^2
  \delta^{ABMN}_{IJKL}
  \\
  {\cal M}_3 \big(v_1^{AB}, S_{3EFGH}, v_4^{MN}\big) \,&=\, \braket{14}^2
  \delta^{ABMN}_{EFGH}
\end{align}
Then we have\footnote{
Here one has used an identity
\begin{align*}\label{}
  \frac12 \delta^{ABMN}_{EFGH}\epsilon_{MN IJKL CD} \,&=\,
  3! \delta^{AB}_{[EF} \epsilon_{GH] IJKL CD}.
\end{align*}
}
\begin{align}\label{}
  {\cal M}_4 \big(v^{AB}, \epsilon\,v_{CD}, S_{EFGH}, S_{IJKL}\big)
  \,=\,
  {1\over\epsilon}
  \bigg( &
  {[21]\over\braket{21}} \delta^{AB}_{CD}
  {\cal M}_3 \big(h_1^-, S_{3EFGH}, S_{4IJKL}\big)
  \nonumber\\
  &+
  {1\over2}  {[23]\over\braket{23}} \epsilon_{MN EFGH CD}
  {\cal M}_3 \big(v_1^{AB}, v_3^{MN}, S_{4IJKL}\big)
  \nonumber\\
  &+
  {1\over2}  {[24]\over\braket{24}} \epsilon_{MN IJKL CD}
  {\cal M}_3 \big(v_1^{AB}, S_{3EFGH}, v_4^{MN}\big)
  \bigg).
\end{align}
By applying straightforwardly the soft gravi-photon theorem \eqref{N=8-soft-gravi_photon-1} to amplitude ${\cal M}_4 \big(v^{AB}, v_{CD}, S_{EFGH}, S_{IJKL}\big)$, one can also obtain the same result.

\subsection*{\it Soft gravi-photino and soft scalar}
According to SUSY Ward identities \eqref{N=8-SWI-gravi_photino-1} and \eqref{N=8-SWI-scalar-1}, one finds
\begin{align}\label{}
  {\cal M}_{n+1}\big(h_1^-, \chi_2^{ABC}, h_3^+, \ldots, h_n^+, \chi_{sDEF}\big) \,&\sim\,
  {\cal O}(\epsilon^0),
  \\
  %
  {\cal M}_{n+1}\big(h_1^-, S_2^{ABCD}, h_3^+, \ldots, h_n^+, S_{sEFGH}\big) \,&\sim\, {\cal O}(\epsilon)
\end{align}
in the holomorphic soft limit $\lambda_s\to\epsilon\lambda_s$.
These results accord with the soft theorems \eqref{N=8-soft-gravi_photino-1} and \eqref{N=8-soft-scalar-1}.

Next we study the {\it 2-scalar 2-gravi-photon amplitude} \eqref{2-scalar-2-gravi-photon-amplitude} which have been discussed previously.
In the holomorphic soft limit $\lambda_4\to\epsilon\lambda_4$, since $\displaystyle \braket{13}^2 [23]^2 = \braket{14}^2 [24]^2 \sim \epsilon^2$, $s \sim t \sim u \sim \epsilon$, this amplitude becomes
\begin{align}\label{}
  {\cal M}_4 (v^- v^+ \phi\phi)\big|_{\lambda_4\to\epsilon\lambda_4} \,&\sim\, \epsilon \,\to\, 0.
\end{align}

In the MHV sector of ${\cal N}=8$ SUGRA, there are also the amplitudes involving four scalars.
Here we analyse a 4-scalar amplitude which was computed in \cite{Kallosh-Lee-Rube-0811.3417}:
\begin{align}\label{}
  {\cal M}_4 \big( & S_{ABCD}, S_{EFGH}, S_{IJKL}, S_{MNPQ}\big) \nonumber\\
  =\,&{tu \over s}\epsilon_{ABCDEFGH}\epsilon_{IJKLMNPQ}
  + {su \over t}\epsilon_{ABCDIJKL}\epsilon_{EFGHMNPQ}
  + {st \over u}\epsilon_{ABCDMNPQ}\epsilon_{EFGHIJKL}
  \nonumber\\
  & + {1 \over 2(4!)^3} \sum_{\sigma} \operatorname{sgn}(\sigma)
  \bigg[
  s\, \epsilon_{1_1 1_2 1_3 1_4 3_1 3_2 4_3 4_4}\, \epsilon_{2_1 2_2 2_3 2_4 4_1 4_2 3_3 3_4}
  + t\, \epsilon_{1_1 1_2 1_3 1_4 2_1 2_2 4_3 4_4}\, \epsilon_{3_1 3_2 3_3 3_4 4_1 4_2 2_3 2_4}
  \nonumber\\
  &\hspace{110pt}
  + u\, \epsilon_{1_1 1_2 1_3 1_4 2_1 2_2 3_3 3_4}\, \epsilon_{4_1 4_2 4_3 4_4 3_1 3_2 2_3 2_4}
  \bigg].
\end{align}
Here $\{ 1_1 1_2 1_3 1_4 \}$ denotes permutations of $\{A,B,C,D\}$ and so forth \cite{Kallosh-Lee-Rube-0811.3417}.
In holomorphic soft limit $\lambda_4\to\epsilon\lambda_4$, since $s \sim t \sim u \sim \epsilon$, the amplitude behaves as
\begin{align}\label{}
  {\cal M}_4 (\phi\phi\phi\phi)\big|_{\lambda_4\to\epsilon\lambda_4} \,&\sim\, \epsilon \,\to\, 0.
\end{align}
A great deal of research shows that amplitudes vanish in soft scalar limit, which indicates a hidden global $E_{7(7)}$ symmetry of classical ${\cal N}=8$ SUGRA \cite{Kallosh-Kugo-0811.3414,He-Zhu-0812.4533, Nima-Cachazo-Kaplan-0808.1446,Bianchi-Elvang-Freedman-0805.0757,Cremmer-1979,Hillmann-0901.1581,Kallosh-1103.4115,Kallosh-Soroush-0802.4106}. This is consist with our soft theorem \eqref{N=8-soft-scalar-1}.

\subsection{SUGRA soft operators as double copy of SYM soft operators}\label{Sec-N=8-double-copy}
As discussed in subsection \ref{Sec-soft-operator-double-copy-1}, the gravity soft operator can be expressed as double copy of gauge theory soft operators.
This also occurs in supersymmetric theories.
In the end of this section, we write the soft operators in ${\cal N}=8$ SUGRA in terms of a sum of some products of soft operators in ${\cal N}=4$ SYM.

First introducing a new operator involving the derivative with respect to Grassmann odd variable $\eta_a^A$ as follows:
\begin{align}\label{soft-operator-seed-eta-1}
  {\frak S}_\eta^1 (s,a) \,&\equiv\, {1\over\braket{s,a}} \eta_s^A{\partial\over\partial\eta_a^A}.
\end{align}
Then the soft operators in ${\cal N}=4$ SYM may be written as
\begin{align}\label{}
  {\cal S}_{\rm SYM}^{(0)} \,&=\, S_{\rm YM}^{(0)} \,=\, {\frak S}^0(x,s,n) + {\frak S}^0(x,s,1),
  \\
  {\cal S}_{\rm SYM}^{(1)} \,&=\, \Big( {\frak S}^1(s,1) - {\frak S}^1(s,n) \Big)
  + \Big( {\frak S}_\eta^1(s,1) - {\frak S}_\eta^1(s,n) \Big).
\end{align}
The `KLT-like formula' of the leading soft factor ${\cal S}^{(0)}$ in gravity has been obtained in section \ref{Sec-soft-operator-double-copy-1}.
The sub-leading soft operator may be written as:
\begin{align}\label{N=8-soft-operator-double-copy-NLO-1}
   {\cal S}^{(1)} \,&=\, \frac12 \sum_{a=1}^{n} s_{sa}
   \Big( {\frak S}^0(x,s,a) + {\frak S}^0(y,s,a) \Big)
   \Big( {\frak S}^1(s,a) + {\frak S}_\eta^1(s,a) \Big)
\end{align}
where $\lambda_x$ and $\lambda_y$ are arbitrary reference spinors.
The sub-sub-leading soft operator may be expressed as
\begin{align}\label{N=8-soft-operator-double-copy-NNLO-1}
   {\cal S}^{(2)} \,&=\, {1\over2}\sum_{a=1}^{n} s_{sa}\,
   \Big( {\frak S}^1(s,a) + {\frak S}_\eta^1(s,a) \Big)
   \Big( {\frak S}^1(s,a) + {\frak S}_\eta^1(s,a) \Big).
\end{align}
As mentioned in subsection \ref{Sec-soft-operator-double-copy-1}, all derivatives in operators only act on amplitudes.
These relations may be derived using the scheme proposed in \cite{Du-Feng-Fu-Wang-1408.4179} by super-KLT relation \cite{Feng-He-1007.0055,Kawai-Lewellen-Tye-1985,
Bjerrum-Bohr-Damgaard-Feng-Sondergaard-1005.4367,
Bjerrum-Bohr-Damgaard-Feng-Sondergaard-1007.3111,
Bjerrum-Bohr-Damgaard-Sondergaard-Vanhove-1010.3933,
Bern-Dixon-Dunbar-Perelstein-Rozowsky-9802162}.

\section{Conclusion and discussions}\label{Sec-Conclusion}
In this work, the super soft theorems were investigated systematically in 4D maximally ${\cal N}=4$ super-Yang-Mills theory and ${\cal N}=8$ supergravity.
We have presented the super soft theorems with rigorous proofs at tree level.
The main results are eq.~\eqref{soft-theorem-N=4-1} for SYM and eq.~\eqref{soft-theorem-N=8-1} for SUGRA.
In ${\cal N}=4$ SYM, several simple examples were examined and the results were in agreement with the super soft theorem exactly.
In ${\cal N}=8$ SUGRA, employing the SUSY Ward identities, we verified the soft gravitino and soft gravi-photon divergences by using leading and sub-leading soft graviton theorem in the MHV sector.
Several four-point amplitudes were also checked in details.

There are several further topics that are fascinating for us.
First, properties of amplitudes involving soft fermion should be investigated more systematically.
In this paper, we discussed the soft gluino divergence for color-ordered amplitudes of ${\cal N}=4$ SYM, the soft gravitino divergence and the soft grav-photino divergence for ${\cal N}=8$ SUGRA amplitudes.
It will be interesting to study the properties of amplitudes involving soft fermion in other theories.

Second, it will be interesting to find other methods to derive the soft theorems.
Let us take an example.
In \cite{Drummond-Henn-0808.2475}, all tree-level superamplitudes in ${\cal N}=4$ SYM were expressed as compact analytical formulas.
By taking the soft limit directly, it gives the soft theorem as shown in appendix \ref{Sec-N=4-alternative}.
The similar formulas for all tree-level superamplitudes in ${\cal N}=8$ SUGRA were also obtained in \cite{Drummond-Spradlin-Volovich-Wen-0901.2363}.
We will also study soft theorem through these formulas in future work.

Finally, more on the relations between the soft theorems and symmetry principle should be understood.
Although the leading and sub-leading soft graviton theorems in gravity \cite{Strominger-1312.2229,He-Lysov-Mitra-Strominger-1401.7026,Kapec-Lysov-Pasterski-Strominger-1406.3312,Strominger-1308.0589,Cachazo-Strominger-1404.4091} and leading and sub-leading soft-photon theorems in massless QED \cite{Lysov-Pasterski-Strominger-1407.3814} were interpreted as symmetries of $\cal S$-matrixes in recent works, very limited information was known for other soft divergences.
Our particular interest is to explore the remarkable relations between super soft theorems and local supersymmetry.

\vskip 10mm
\section*{Acknowledgments}
The author would like to thank Professor Jun-Bao Wu for suggesting this project as well as numerous discussions and valuable comments through out all the stages of the work.
He would also especially like to thank Professor Chuan-Jie Zhu for his support, guidance, discussions and careful reading of the manuscript.
He is also grateful to Da-Ping Liu, Wen-Jian Pan, Yu Tian, Gang Yang and Hongbao Zhang for various helpful discussions.
This work was supported by the National Natural Science Foundation of China under Grants No.~11135006.

\newpage
\appendix
\section{Alternative derivation of soft theorem in ${\cal N}=4$ SYM}\label{Sec-N=4-alternative}
In this appendix, we rederive the soft theorem of ${\cal N}=4$ SYM by using formulaes for all tree-level superamplitudes which were given in \cite{Drummond-Henn-0808.2475}.

First of all, we have to summarize briefly main results of Drummond and Henn's paper \cite{Drummond-Henn-0808.2475}.
Three-point amplitudes are fundamental in BCFW-construction of higher-point amplitudes.
In section \ref{Sec-N=4}, three-point MHV and Googly (or anti-MHV) superamplitudes of ${\cal N}=4$ SYM have been presented.
By solving super-BCFW resursion, it is easy to obtain general $n$-point $(n>3)$ MHV superamplitudes
\begin{align}\label{}
  {\mathscr A}_n^{\rm MHV} \big( \{\lambda_a,\tilde\lambda_a,\eta_a\} \big) \,&=\,
  {\delta^4(p) \delta^8(q) \over \braket{12}\braket{23}\cdots\braket{n1} }.
\end{align}
The $n$-point MHV superamplitude is simple and compact, just as Parke-Taylor formula of the pure gluonic amplitude.
The delta functions $\delta^4(p)$ and $\delta^8(q)$ are consequences of translation invariance and supersymmetry.
Therefore all tree-level superamplitudes, not just MHV, contain delta function factor $\displaystyle \delta^4(p) \delta^8(q)$ in ${\cal N}=4$ SYM.
So it is very convenient to factor out the MHV superamplitude,
\begin{align}\label{}
  {\mathscr A}_n \,&=\, {\mathscr A}_n^{\rm MHV}\,{\cal P}_n.
\end{align}
Here ${\cal P}_n$ is a function of spinors $\lambda_a$, $\tilde\lambda_a$ and Grassmann variables $\eta_a^A$ and one can express this quantity as following form:
\begin{align}\label{}
  {\cal P}_n \,&=\, {\cal P}_n^{\rm MHV} + {\cal P}_n^{\rm NMHV} + \cdots + {\cal P}_n^{\overline{\rm MHV}}.
\end{align}
Of course ${\cal P}_n^{\rm MHV}=1$, and the N$^k$MHV function $\displaystyle {\cal P}_n^{\rm N^{\it k}MHV}$ has Grassmann degree $4k$.

Turning to the NMHV sector, the function ${\cal P}_n^{\rm NMHV}$ is given by \cite{Drummond-Henn-0808.2475}
\begin{align}\label{P-factor-n-pt-1}
  {\cal P}_n^{\rm NMHV} \,=\, \sum_{2\leq a < b \leq n-1} R_{n;ab}.
\end{align}
Here $R_{n;ab}$ is a dual superconformal invariant \cite{Bianchi-Elvang-Freedman-0805.0757,
Elvang-Freedman-Kiermaier-0911.3169,
Drummond-Spradlin-Volovich-Wen-0901.2363,
Drummond-Henn-Plefka-0902.2987,
Drummond-Henn-Korchemsky-Sokatchev-0807.1095,
Brandhuber-Heslop-Travaglini-0807.4097,
Korchemsky-Sokatchev-0906.1737,
Korchemsky-Sokatchev-1002.4625,
Drummond-Henn-0808.2475,
Beisert-1012.3982}
\begin{align}\label{R-invariant-nab-1}
  R_{n;ab} \,&=\, {
  \braket{a, a\!-\!1}\braket{b, b\!-\!1} \delta^4 \big(\Xi_{n;ab}\big)
  \over
  x_{ab}^2 \braket{n | x_{na}x_{ab} | b} \braket{n | x_{na}x_{ab} | b\!-\!1}
  \braket{n | x_{nb}x_{ba} | a} \braket{n | x_{nb}x_{ba} | a\!-\!1}
  }
\end{align}
where
\begin{align}\label{}
  x_{ij} \,&\equiv\, k_i + k_{i+1} + \cdots + k_{j-1},  \\
  \theta_{ij} \,&\equiv\, q_i + q_{i+1} + \cdots + q_{j-1},  \qquad q_a\equiv\lambda_a\eta_a,
\end{align}
and the Grassmann odd quantity $\Xi_{n;ab}$ is defined by
\begin{align}\label{}
  \Xi_{n;ab} \,&\equiv\, \Braket{n | x_{na}x_{ab} | \theta_{bn}} + \Braket{n | x_{nb}x_{ba} | \theta_{an}}
  \nonumber\\
  &=\, \bra{n}\Bigg( x_{na} x_{ab} \sum_{i=b}^{n-1} \ket{i}\eta_i +
  x_{nb} x_{ba} \sum_{i=a}^{n-1}\ket{i}\eta_i \Bigg).
\end{align}
Obviously, this quantity is {\it independent} of $\eta_1$ and $\eta_n$.
In fact, it is relevant to a special gauge choice.

Similarly, NNMHV function also may be constructed as follows
\begin{align}\label{P-factor-NNMHV-1}
  {\cal P}_n^{\rm NNMHV} \,&=\, \sum_{2 \leq a_1, b_1 \leq n-1} R^{0;0}_{n,a_1b_1}
  \Bigg(
  \sum_{a_1+1 \leq a_2, b_2 \leq b_1} R^{0; a_1 b_1}_{n; b_1a_1; a_2b_2}
  + \sum_{b_1 \leq a_2, b_2 \leq n-1} R^{a_1b_1; 0}_{n; a_2b_2}
  \Bigg).
\end{align}
Here has involved a quantity which is a generalization of the $R$-invariant,
\begin{align}\label{R-invariant-gene-1}
  R_{n; b_1 a_1; b_2 a_2; \ldots; b_r a_r; ab} \,&=\, {
  \braket{a, a-1}\braket{b, b-1} \delta^4 \big(\Xi_{n; b_1a_1; b_2a_2; \ldots; b_r a_r; ab}\big) \over
  x_{ab}^2 \bra{\xi} x_{a_r a} x_{ab} \ket{b}  \bra{\xi} x_{a_r a} x_{ab} \ket{b-1}
  \bra{\xi} x_{a_r b} x_{ba} \ket{a}  \bra{\xi} x_{a_r b} x_{ba} \ket{a-1}
  }.
\end{align}
For details, see \cite{Drummond-Henn-0808.2475}.

More generally, all the $\displaystyle {\cal P}_n^{\rm N^\text{$k$}MHV}$ functions can be written in terms of the quantities $R_{n; b_1 a_1; b_2 a_2; \ldots; b_r a_r; ab}$.
It is somewhat surprising that all quantities \eqref{R-invariant-gene-1} are independent of $\eta_1$ and $\eta_n$ and so are all $\displaystyle {\cal P}_n^{\rm N^\text{$k$}MHV}$, i.e.,
\begin{align}\label{}
  {\partial \over \partial\eta_1^A} {\cal P}_n \,=\, {\partial \over \partial\eta_n^A} {\cal P}_n
  \,=\, 0.
\end{align}
In fact, this reflects the special gauge choice of shifted legs in BCFW recursion.

Now we turn to study soft theorem.
The $n$-point MHV superamplitude is not only the simplest in all $n$-point amplitudes, but a common factor for all amplitudes.
So we begin with MHV sectors.

First we write the delta function $\delta^8(q)$ as
\begin{align}\label{}
  \delta^8(q) \,=\, \braket{1,n}^4\,
  \delta^4\left( {\braket{n,s}\over\braket{n,1}}\eta_s^A +
  \eta_1^A + \sum_{a=2}^{n-1} {\braket{n,a}\over\braket{n,1}}\eta_a^A \right)
  \delta^4\left( {\braket{1,s}\over\braket{1,n}}\eta_s^B +
  \eta_n^B + \sum_{a=2}^{n-1} {\braket{1,a}\over\braket{1,n}}\eta_a^B \right).
\end{align}
When the soft particle is gluon, we have
\begin{align}
  {\mathscr A}_{n+1}^{\rm MHV} \big(\Phi_1, \ldots, \Phi_n, g_s^+\big) \,&=\,
  {\delta^4(p)\, \braket{1,n}^4 \over \braket{s,1}\braket{1,2}\cdots\braket{n,s} }
  \delta^4\left( \eta_1^A + \sum_{a=2}^{n-1} {\braket{n,a}\over\braket{n,1}}\eta_a^A \right)
  \delta^4\left( \eta_n^B + \sum_{a=2}^{n-1} {\braket{1,a}\over\braket{1,n}}\eta_a^B \right)
  \nonumber\\
  &=\,
  {\braket{n,1} \over \braket{s,1}\braket{n,s} }
  {\delta^4(p) \over \braket{1,2}\cdots\braket{n,1} }
  \delta^8\left( \sum\nolimits_{a=1}^n \lambda_a^\alpha\eta_a^A\right).
  \label{soft-gluon-rescal-1}
\end{align}
In the holomorphic soft limit $\lambda_s\to\epsilon\lambda_s$, it can give the leading soft factor of Yang-Mills amplitude.

When the soft particle is gluino, we have
\begin{align}\label{}
  \eta_s^A {\mathscr A}_{n+1}^{\rm MHV} & \big(\Phi_1, \ldots, \Phi_n, \psi_{sA}\big) \,=\,
  {\braket{n,1} \over \braket{n,s}\braket{s,1} }
  {\delta^4(p)\, \braket{1,n}^4 \over \braket{1,2}\cdots\braket{n,1} }
  \nonumber\\
  &\times\Bigg\{
  {\braket{n,s}\over\braket{n,1}} \times (-1)^{A+1}\eta_s^A
  \prod_{B \ne A} \bigg( \eta_1^B + \sum_{a=2}^{n-1} {\braket{n,a}\over\braket{n,1}}\eta_a^B \bigg)
  \prod_{C} \bigg( \eta_n^C + \sum_{a=2}^{n-1} {\braket{1,a}\over\braket{1,n}}\eta_a^C \bigg)
  \nonumber\\
  &~~~+{\braket{1,s}\over\braket{1,n}} \times (-1)^{A+1}\eta_s^A
  \prod_{C} \bigg( \eta_1^C + \sum_{a=2}^{n-1} {\braket{n,a}\over\braket{n,1}}\eta_a^C \bigg)
  \prod_{B \ne A} \bigg( \eta_n^B + \sum_{a=2}^{n-1} {\braket{1,a}\over\braket{1,n}}\eta_a^B \bigg)
  \Bigg\}
  \nonumber\\
  =~&\,
  {\braket{n,1} \over \braket{n,s}\braket{s,1} }
  \Bigg(
  {\braket{n,s}\over\braket{n,1}} \eta_s^A{\partial \over \partial\eta_1^A}
  + {\braket{1,s}\over\braket{1,n}} \eta_s^A{\partial \over \partial\eta_n^A}
  \Bigg)
  {\mathscr A}_{n}^{\rm MHV} \big(\Phi_1, \ldots, \Phi_n \big).
  \label{soft-gluino-rescal-1}
\end{align}
In the holomorphic soft limit $\lambda_s\to\epsilon\lambda_s$, it gives ${\cal O}(\epsilon^{-1})$ order soft divergence.
When the soft particle is scalar, it is also easy to see that MHV superamplitudes are invariant under the holomorphic soft rescaling $\lambda_s\to\epsilon\lambda_s$.

Another task is to see the function ${\cal P}_{n+1}$ in the holomorphic soft limit.
First we consider soft behavior of NMHV function\footnote{Here we fix gauge such that the function ${\cal P}^{\rm NMHV}_{n+1}$ is independent of $\eta_n$  and $\eta_s$.}:
\begin{align}\label{P-factor-n+1-pt-1}
  {\cal P}^{\rm NMHV}_{n+1} (1, \ldots, \hat n, \hat s) \,&=\, \sum_{1 \leq a < b \leq n-1}
  R_{n;ab}(1, \ldots, \hat n, \hat s)
\end{align}
Here dual superconformal invariant $R_{n;ab}$ is given by
\begin{align}\label{R-invariant-nab-2}
  R_{n;ab}(1, \ldots, \hat n, \hat s) \,&=\, {
  \braket{a, a\!-\!1}\braket{b, b\!-\!1} \delta^4 \big(\Xi_{n;ab}\big)
  \over
  \tilde x_{ab}^2 \braket{n | \tilde x_{na}\tilde x_{ab} | b} \braket{n | \tilde x_{na}\tilde x_{ab} | b\!-\!1}
  \braket{n | \tilde x_{nb}\tilde x_{ba} | a} \braket{n | \tilde x_{nb}\tilde x_{ba} | b\!-\!1}
  }
\end{align}
where
\begin{align}\label{}
  \Xi_{n;ab} \,&=\,
  \bra{n}\Bigg( \tilde x_{na} \tilde x_{ab} \sum_{i=b}^{n-1} \ket{i}\eta_i +
  \tilde x_{nb} \tilde x_{ba} \sum_{i=a}^{n-1}\ket{i}\eta_i \Bigg).
\end{align}
Here the dual variable $\tilde x_{ab}$ corresponds to the ordering $(1, \ldots, n, s)$ of $(n+1)$-point {\it color-ordered} superamplitude ${\mathscr A}_{n+1}(\Phi_1, \ldots, \Phi_n, \Phi_s)$ while the variable $x_{ab}$ corresponds to the ordering $(1, \ldots, n)$ of $n$-point superamplitude ${\mathscr A}_{n}(\Phi_1, \ldots, \Phi_n)$.

In the holomorphic soft limit $\lambda_s\to\epsilon\lambda_s$,
\begin{align}\label{}
  \braket{a, a\!-\!1} \,&=\,
  \left\{ \begin{array}{ll}
     \epsilon\braket{1,s} & ~~\text{for $a=1$},\\
     \braket{a, a\!-\!1}  & ~~\textrm{for $a\geq 2$},
  \end{array} \right.
  \\
  \tilde x_{na} \,&=\, \epsilon k_s + k_1 + \cdots + k_{a-1} \,=\, x_{na} + \epsilon k_s,
  \\
  \tilde x_{ab} \,&=\, x_{ab}  ~~~\text{for}~1 \leq a < b \leq n-1,
  \\
  \tilde x_{ba} \,&=\, x_{ba} + \epsilon k_s  ~~~\text{for}~1 \leq a < b \leq n-1,
  \\
  {1 \over \tilde x_{na}} \,&=\, {1 \over x_{na}} + {\cal O}(\epsilon),
  \\
  {1 \over \tilde x_{ba}} \,&=\, {1 \over x_{ba}} + {\cal O}(\epsilon) ~~~\text{for}~1 \leq a < b \leq n-1.
\end{align}
Substituting above formulas into eq.~\eqref{P-factor-n+1-pt-1}, one finds
\begin{align}\label{P-factor-MHV-soft-limit-1}
  {\cal P}^{\rm NMHV}_{n+1} (1, \ldots, \hat n, \hat s) \,&=\,
  \sum_{2 \leq a < b \leq n-1}
  R_{n;ab}(\hat 1, \ldots, \hat n)  \,+\, {\cal O}(\epsilon) \,=\,
  {\cal P}^{\rm NMHV}_{n} (\hat 1, \ldots, \hat n) \,+\, {\cal O}(\epsilon).
\end{align}
in the holomorphic soft limit $\lambda_s\to\epsilon\lambda_s$.
Applying same analysis to ${\cal P}_n^{\rm NNMHV}$, eq.~\eqref{P-factor-NNMHV-1}, one also obtains
\begin{align}\label{P-factor-MHV-soft-limit-1}
  {\cal P}^{\rm NNMHV}_{n+1} (1, \ldots, \hat n, \hat s) \,&=\,
  {\cal P}^{\rm NNMHV}_{n} (\hat 1, \ldots, \hat n ) \,+\, {\cal O}(\epsilon).
\end{align}
Using the same method, one can show that the similar conclusion holds for all N${}^k$MHV sectors of ${\cal N}=4$ SYM, i.e.,
\begin{align}\label{P-factor-soft-X-1}
  {\cal P}_{n+1}\big(1,\ldots, n, s\big) \,&=\,  {\cal P}_{n}\big(1, \ldots, n\big) \,+\, {\cal O}(\epsilon).
\end{align}

Let us notice that the function $\displaystyle {\cal P}_{n+1}$ is independent of Grassmann variables $\eta_s^A$ with a certain gauge choice.
Using eq.~\eqref{P-factor-soft-X-1} and eq.~\eqref{soft-gluon-rescal-1}, it is easy to obtain the leading soft gluon divergence,
\begin{align}
  {\mathscr A}_{n+1} \big(\ldots, g_s^+\big) \,\equiv\,
  {\mathscr A}_{n+1}^{\rm MHV} \big(\ldots, g_s^+\big)  {\cal P}_{n+1}
  \,=\,
  {1\over\epsilon^2}  {\braket{n,1} \over \braket{s,1}\braket{n,s} }
  {\mathscr A}_{n}^{\rm MHV}
  \Big( {\cal P}_{n} \,+\, {\cal O}(\epsilon) \Big).
\end{align}
in the holomorphic soft limit $\lambda_s\to\epsilon\lambda_s$.
As was discussed earlier, there is no sub-leading soft gluon divergence and here we no longer discuss it.
For soft gluino, by using eq.~\eqref{P-factor-soft-X-1} and eq.~\eqref{soft-gluino-rescal-1}, one gets
\begin{align}\label{}
  {\mathscr A}_{n+1} \big(\ldots, \psi_{sA}\big) \,&=\,
  {1\over\epsilon}  {\braket{n,1} \over \braket{n,s}\braket{s,1} }
  \Bigg(
  {\braket{n,s}\over\braket{n,1}} {\partial \over \partial\eta_1^A}
  + {\braket{1,s}\over\braket{1,n}} {\partial \over \partial\eta_n^A}
  \Bigg)
  {\mathscr A}_{n}^{\rm MHV} \Big( {\cal P}_{n}  \,+\, {\cal O}(\epsilon) \Big).
\end{align}
in the holomorphic soft limit $\lambda_s\to\epsilon\lambda_s$.
Both ${\mathscr A}_{n+1}^{\rm MHV}$ and ${\cal P}_{n+1}$ have no singular term in the holomorphic soft limit $\lambda_s\to\epsilon\lambda_s$ of a scalar.
This implies that there exists no singularity when an external scalar becomes soft in an on-shell amplitude.

\newpage
\section{Sub-sub-leading soft operator in ${\cal N}=8$ SUGRA}\label{Sec-N=8-NNLO-Soft-Operators}
In this appendix, we compute the sub-sub-leading soft operator in ${\cal N}=8$ SUGRA by a Taylor expansion in detail.

As shown in section \ref{Sec-N=8}, in the holomorphic soft limit $\lambda_s \to \epsilon\lambda_s$, the superamplitude ${\mathscr M}_{n+1}$ becomes
\begin{align}\label{N=8-sub-sub-soft-operator-X-1}
  {\mathscr M}_{n+1} (\epsilon) &=
  {1\over \epsilon^3}\sum_{a=1}^{n-1} {[s,a] \braket{n,a}^2 \over \braket{s,a} \braket{n,s}^2 }
  \nonumber\\
  &\times
  {\mathscr M}_n \Big(
  \ldots,
  \{\lambda_a, \tilde\lambda_a + \epsilon {\braket{n,s} \over \braket{n,a}}\tilde\lambda_s,
  \eta_a + \epsilon {\braket{n,s} \over \braket{n,a}} \eta_s \}, \ldots,
  \{\lambda_n, \tilde\lambda_n + \epsilon {\braket{s,a} \over \braket{n,a}}\tilde\lambda_s,
  \eta_n + \epsilon {\braket{s,a} \over \braket{n,a}}\eta_s \} \Big).
\end{align}
Let us denote
\begin{align}\label{}
  G(\epsilon) \,&=\, {\mathscr M}_n \Big(
  \ldots,
  \{\lambda_a, \tilde\lambda_a + \epsilon {\braket{n,s} \over \braket{n,a}}\tilde\lambda_s,
  \eta_a + \epsilon {\braket{n,s} \over \braket{n,a}} \eta_s \}, \ldots,
  \{\lambda_n, \tilde\lambda_n + \epsilon {\braket{s,a} \over \braket{n,a}}\tilde\lambda_s,
  \eta_n + \epsilon {\braket{s,a} \over \braket{n,a}}\eta_s \} \Big)
\end{align}
then expand it in infinitesimal soft parameter $\epsilon$:
\begin{align}\label{}
  G(\epsilon) \,&=\, G(0) + \epsilon G'(0) + \frac12\epsilon^2 G''(0) + {\cal O}(\epsilon^3).
\end{align}
The first two orders contribute to the leading and the sub-leading soft operators respectively.
The second order in $\epsilon$ is
\begin{align}
  \label{soft-N=8-exapn-NNLO-1}
  G''(0) \,&=\, G_{\lambda}''(0) + G_{\lambda\eta}''(0) + G_{\eta}''(0), \\
  \label{soft-N=8-exapn-NNLO-1-1}
  G_{\lambda}''(0) \,&=\,
  \tilde\lambda_{s\dot\alpha} \tilde\lambda_{s\dot\beta}
  \Bigg(
  {\braket{n,s}^2\over\braket{a,n}^2}
  {\partial^2\over\partial\tilde\lambda_{a\dot\alpha} \partial\tilde\lambda_{a\dot\beta}}
  -2 {\braket{n,s}\braket{a,s}\over\braket{n,a}^2}
  {\partial^2\over\partial\tilde\lambda_{a\dot\alpha} \partial\tilde\lambda_{n\dot\beta}}
  + {\braket{a,s}^2\over\braket{a,n}^2}
  {\partial^2\over\partial\tilde\lambda_{n\dot\alpha} \partial\tilde\lambda_{n\dot\beta}}
  \Bigg)
  {\mathscr M}_n, \\
  \label{soft-N=8-exapn-NNLO-1-2}
  G_{\lambda\eta}''(0) \,&=\,
  2\tilde\lambda_{s\dot\alpha} \eta_s^A \Bigg(
  {\braket{n,s}^2\over\braket{a,n}^2}
  {\partial^2\over\partial\tilde\lambda_{a\dot\alpha} \partial\eta_a^A}
  + {\braket{n,s}\braket{s,a}\over\braket{a,n}^2}
  \bigg(
  {\partial^2\over\partial\tilde\lambda_{a\dot\alpha} \partial\eta_n^A}
  + {\partial^2\over\partial\tilde\lambda_{n\dot\alpha} \partial\eta_a^A}
  \bigg)
  + {\braket{s,a}^2\over\braket{n,a}^2}
  {\partial^2\over\partial\tilde\lambda_{n\dot\alpha} \partial\eta_n^A}
  \Bigg)
  {\mathscr M}_n,  \\
  \label{soft-N=8-exapn-NNLO-1-3}
  G_{\eta}''(0) \,&=\,
  \eta_s^A \eta_s^B
  \Bigg(
  {\braket{n,s}^2\over\braket{a,n}^2}
  {\partial^2\over\partial\eta_a^B \partial\eta_a^A}
  -2 {\braket{n,s}\braket{a,s}\over\braket{n,a}^2}
  {\partial^2\over\partial\eta_a^B \partial\eta_n^A}
  + {\braket{a,s}^2\over\braket{a,n}^2}
  {\partial^2\over\partial\eta_n^B \partial\eta_n^A}
  \Bigg)
  {\mathscr M}_n.
\end{align}
It is important to notice that here the derivative with respect to the Grassmann variable $\eta^A$ is {\it left derivative}.
Plugging the second derivative $G''(0)$ on the right hand side of eq.~\eqref{N=8-sub-sub-soft-operator-X-1}, we obtain
\begin{align}\label{}
  {\cal S}^{(2)}{\cal M}_n \,&=\, {1\over 2\epsilon}\sum_{a=1}^{n-1}
  {[s,a] \braket{n, a}^2 \over \braket{s,a}\braket{n, s}^2}\,
  \Big(G_{\lambda}''(0) + G_{\lambda\eta}''(0) + G_{\eta}''(0) \Big).
  \label{NNLO-soft-factor-N=8-1}
\end{align}

The first term \eqref{soft-N=8-exapn-NNLO-1-1} on the right hand side of eq.~\eqref{NNLO-soft-factor-N=8-1}
has obtained in Cachazo-Strominger's paper \cite{Cachazo-Strominger-1404.4091}:
\begin{align}\label{}
  S^{(2)} \,&=\, {1\over 2} \sum_{a=1}^{n} {[s,a] \over \braket{s,a}}
  \tilde\lambda_{s\dot\alpha}\tilde\lambda_{s\dot\beta}
  {\partial^2\over\partial\tilde\lambda_{a\dot\alpha}\partial\tilde\lambda_{a\dot\beta}}.
\end{align}
By using similar derivation, it is easy to write down the third term \eqref{soft-N=8-exapn-NNLO-1-3} on the right hand side of eq.~\eqref{NNLO-soft-factor-N=8-1}:
\begin{align}\label{}
  {\cal S}_\eta^{(2)} \,&=\, {1\over 2} \sum_{a=1}^{n} {[s,a] \over \braket{s,a}}
  \eta_s^A\eta_s^B {\partial^2\over\partial\eta_a^B\partial\eta_a^A}.
\end{align}

The second term \eqref{soft-N=8-exapn-NNLO-1-2} on the right hand side of eq.~\eqref{NNLO-soft-factor-N=8-1} is given by
\begin{align}\label{}
  {\cal S}_{\lambda\eta}^{(2)}{\mathscr M}_n \,&=\, \sum_{a=1}^{n-1}
  \tilde\lambda_{s\dot\alpha} \eta_s^A \Bigg(
  {[s,a] \over \braket{s,a}}
  {\partial^2\over\partial\tilde\lambda_{a\dot\alpha} \partial\eta_a^A}
  + {[s,a] \over \braket{n, s}}
  \bigg(
  {\partial^2\over\partial\tilde\lambda_{a\dot\alpha} \partial\eta_n^A}
  + {\partial^2\over\partial\tilde\lambda_{n\dot\alpha} \partial\eta_a^A}
  \bigg)
  + {\braket{s,a} [s,a]  \over \braket{n, s}^2}
  {\partial^2\over\partial\tilde\lambda_{n\dot\alpha} \partial\eta_n^A}
  \Bigg)
  {\mathscr M}_n.
  \label{soft-oper-N=8-2nd-1}
\end{align}
Noticing that
\begin{align}\label{}
  \sum_{a=1}^{n-1} \braket{s,a} [s,a] \,=\, \sum_{a=1}^{n-1} 2k_s\cdot k_a
  \,=\, -2k_s\cdot \big(k_n + k_s\big) \,=\, -2k_s\cdot k_n
  \,=\, -\braket{s,n}[s,n],
  \label{useful-id-1}
\end{align}
then the last term of eq.~\eqref{soft-oper-N=8-2nd-1}  becomes
\begin{align}\label{}
  \sum_{a=1}^{n-1}
  {\braket{s,a} [s,a]  \over \braket{n, s}^2} \tilde\lambda_{s\dot\alpha} \eta_s^A
  {\partial^2\over\partial\tilde\lambda_{n\dot\alpha} \partial\eta_n^A}
  {\mathscr M}_n
  \,=\,
  - {[s,n]  \over \braket{s,n}} \tilde\lambda_{s\dot\alpha} \eta_s^A
  {\partial^2\over\partial\tilde\lambda_{n\dot\alpha} \partial\eta_n^A}
  {\mathscr M}_n.
\end{align}
Write the second term of eq.~\eqref{soft-oper-N=8-2nd-1} as follows:
\begin{align}\label{}
  \sum_{a=1}^{n-1} {[s,a] \over \braket{n, s}}
  \tilde\lambda_{s\dot\alpha} \eta_s^A
  {\partial^2\over\partial\tilde\lambda_{a\dot\alpha} \partial\eta_n^A}
  {\mathscr M}_n
  \,&=\, {\eta_s^A \tilde\lambda_{s\dot\beta} \tilde\lambda_{s\dot\alpha} \over \braket{n, s}}
  \sum_{a=1}^{n-1} \tilde\lambda_a^{\dot\alpha}
  {\partial^2\over\partial\tilde\lambda_{a\dot\beta}\, \partial\eta_n^A}
  {\mathscr M}_n.
\end{align}
Using the global angular momentum conservation, we get
\begin{align}\label{}
  \tilde\lambda_{s\dot\beta}  \tilde\lambda_{s\dot\alpha}
  \sum_{a=1}^{n-1} \tilde\lambda_a^{\dot\alpha}
  {\partial\over\partial\tilde\lambda_{a\dot\beta}}
  {\mathscr M}_n
  \,=\, -
  \tilde\lambda_{s\dot\beta}  \tilde\lambda_{s\dot\alpha}
  \tilde\lambda_n^{\dot\alpha} {\partial\over\partial\tilde\lambda_{n\dot\beta}}
  {\mathscr M}_n
  \,=\, - [s,n] \tilde\lambda_{s\dot\alpha} {\partial\over\partial\tilde\lambda_{n\dot\alpha}}
  {\mathscr M}_n.
\end{align}
Thus we have
\begin{align}\label{}
  \sum_{a=1}^{n-1} {[s,a] \over \braket{n, s}}
  \tilde\lambda_{s\dot\alpha} \eta_s^A
  {\partial^2\over\partial\tilde\lambda_{a\dot\alpha} \partial\eta_n^A}
  {\mathscr M}_n
  \,&=\, {[s,n] \over \braket{s,n}} \tilde\lambda_{s\dot\alpha} \eta_s^A {\partial^2\over\partial\tilde\lambda_{n\dot\alpha}\,\partial\eta_n^A}
  {\mathscr M}_n.
\end{align}
The third term of eq.~\eqref{soft-oper-N=8-2nd-1} may be written as:
\begin{align}\label{}
  \sum_{a=1}^{n-1} {[s,a] \over \braket{n, s}}
  \tilde\lambda_{s\dot\beta} \eta_s^A
  {\partial^2\over\partial\tilde\lambda_{n\dot\beta} \partial\eta_a^A}
  {\mathscr M}_n
  \,&=\, {\eta_{s}^A \tilde\lambda_{s\dot\beta} \tilde\lambda_{s\dot\alpha} \over \braket{n, s}}
  \sum_{a=1}^{n-1}  \tilde\lambda_a^{\dot\alpha}\,
  {\partial^2\over\partial\eta_a^A\, \partial\tilde\lambda_{n\dot\beta}}
  {\mathscr M}_n.
\end{align}
Using the supersymmetry\footnote{
The SUSY requires that
\begin{align*}
  \sum_{a=1}^n \lambda_a^\alpha \eta_a^A {\mathscr M}_n \,=\,
  \sum_{a=1}^n \tilde\lambda_a^{\dot\alpha} {\partial\over\partial\eta_a^A} {\mathscr M}_n \,=\,0
\end{align*}
for $n$-point on-shell superamplitudes.
}, we have
\begin{align}\label{}
  \tilde\lambda_{s\dot\alpha}\sum_{a=1}^{n-1} \tilde\lambda_a^{\dot\alpha}
  {\partial\over\partial\eta_a^A}
  {\mathscr M}_n
  \,=\, - \tilde\lambda_{s\dot\alpha}
  \tilde\lambda_n^{\dot\alpha} {\partial\over\partial\eta_n^A}
  {\mathscr M}_n
  \,=\, -[s,n] {\partial\over\partial\eta_n^A}
  {\mathscr M}_n.
\end{align}
Thus we obtain the third term of eq.~\eqref{soft-oper-N=8-2nd-1}:
\begin{align}\label{}
  \sum_{a=1}^{n-1} {[s,a] \over \braket{n, s}}
  \tilde\lambda_{s\dot\beta} \eta_s^A
  {\partial^2\over\partial\tilde\lambda_{n\dot\beta} \partial\eta_a^A}
  {\mathscr M}_n
  \,&=\, {[s,n] \over \braket{s,n}}
  \tilde\lambda_{s\dot\alpha} \eta_s^A
  {\partial^2\over \partial\tilde\lambda_{n\dot\alpha} \partial\eta_n^A}
  {\mathscr M}_n.
\end{align}
Combining all the contributions, we get
\begin{align}\label{}
  {\cal S}_{\lambda\eta}^{(2)}{\mathscr M}_n \,&=\, \sum_{a=1}^{n} {[s,a] \over \braket{s,a}} \tilde\lambda_{s\dot\alpha} \eta_s^A
  {\partial^2\over\partial\tilde\lambda_{a\dot\alpha} \partial\eta_a^A} {\mathscr M}_n.
\end{align}

Finally, we obtain a complete sub-sub-leading soft operator in ${\cal N}=8$ SUGRA
\begin{align}\label{}
  {\cal S}^{(2)} \,&=\, \frac12 \sum_{a=1}^{n} {[s,a] \over \braket{s,a}}
  \left(
  \tilde\lambda_{s\dot\alpha} \tilde\lambda_{s\dot\beta}
  {\partial^2\over\partial\tilde\lambda_{a\dot\alpha} \partial\tilde\lambda_{a\dot\beta}}
  + 2 \tilde\lambda_{s\dot\alpha} \eta_s^A
  {\partial^2\over\partial\tilde\lambda_{a\dot\alpha} \partial\eta_a^A}
  + \eta_s^A\eta_s^B {\partial^2\over\partial\eta_a^B \partial\eta_a^A}
  \right).
\end{align}

\end{document}